\documentclass{jpp}
\usepackage{graphicx}

\usepackage[utf8]{inputenc}
\usepackage[T1]{fontenc}
\usepackage{amsmath}
\usepackage{amssymb}
\usepackage{babel}
\usepackage{authblk}
\usepackage{graphicx}

\shorttitle{Non-equilibrium $\alpha$-effect from cross-helicity}
\shortauthor{K. A. Mizerski, N. Yokoi and A. Brandenburg}

\title{
Cross-helicity effect on $\alpha$-type dynamo\\ in non-equilibrium turbulence
}

\author[1]{Krzysztof A. Mizerski\corresp{\email{kamiz@igf.edu.pl}}}
\author[2]{Nobumitsu Yokoi}
\author[3,4,5,6]{Axel Brandenburg}
\affil[1]{{\small{Department of Magnetism, Institute of Geophysics, Polish Academy of Sciences, Ksiecia Janusza 64,
01-452 Warsaw, Poland}}}
\affil[2]{{\small{Institute of Industrial Science, University of Tokyo, Komaba, Meguro, Tokyo 153-8505, Japan}}}
\affil[3]{{\small{Nordita, KTH Royal Institute of Technology and Stockholm University, Hannes Alfv\'ens v\"ag 12, 10691 Stockholm, Sweden}}}
\affil[4]{{\small{The Oskar Klein Centre, Department of Astronomy, Stockholm University, AlbaNova, 10691 Stockholm, Sweden}}}
\affil[5]{{\small{School of Natural Sciences and Medicine, Ilia State University, 0194 Tbilisi, Georgia}}}
\affil[6]{{\small{McWilliams Center for Cosmology and Department of Physics, Carnegie Mellon University, Pittsburgh, Pennsylvania 15213, USA}}}

\begin{document}

\maketitle

\begin{abstract}
Turbulence is typically not in equilibrium, i.e.\ mean quantities such as the mean energy and helicity are typically time-dependent. The effect of non-stationarity on the turbulent hydromagnetic dynamo process is studied here with the use of the two-scale direct-interaction approximation (TSDIA), which allows to explicitly relate the mean turbulent Reynolds and Maxwell stresses and the mean electromotive force (EMF) to the spectral characteristics of turbulence, such as e.g.\ the mean energy, as well as kinetic and cross-helicity. It is demonstrated, that the non-equilibrium effects can enhance the dynamo process when the magnetohydrodynamic (MHD) turbulence is both helical and cross-helical. This effect is based on the turbulent infinitesimal-impulse cross-response functions, which do not affect turbulent flows in equilibrium. The evolution and sources of the cross-helicity in MHD turbulence is also discussed.
\end{abstract}


\section{Introduction}
The effect of hydromagnetic dynamo action is ubiquitous in astrophysical plasmas e.g.\ in stellar and planetary interiors, accretion discs or the interstellar medium (cf.\ Roberts and Soward 1972, Brandenburg and Subramanian 2005, Dormy and Soward 2007, Roberts and King 2013, Balbus and Hawley 1991a,b). This is particularly important in view of the recent advancement of tokamak devices, reaching very high plasma temperatures, thus giving hope for the production of thermonuclear fusion power (cf.\ Li \emph{et al.} 2019, Gibney 2022). The investigations of the large-scale dynamo mechanisms in magnetohydrodynamic (MHD) turbulence, that is those that lead to generation of large-scale magnetic fields, is mainly limited to equilibrium, i.e.\ statistically stationary turbulence.  

One of the widely known and often invoked mechanisms is the so-called $\alpha$-effect, which requires chirality (lack of reflexional symmetry) in the turbulent flow, and this requires some mechanism that breaks the `up-down' symmetry of the system, cf.\ Krause and R\"adler (1980), Dormy and Soward (2007), Moffatt and Dormy (2019). A large-scale electromotive force (EMF) is then generated and this leads to the amplification of magnetic energy. The lack of reflectional symmetry is typically introduced by stratification and background rotation and a useful measure of the flow chirality is the kinetic helicity, $\langle\mathbf{U}\cdot\nabla\times\mathbf{U}\rangle$, where $\langle\cdot\rangle$ denotes the ensemble mean. Another pseudoscalar quantity of importance in dynamo theory is the cross-helicity $\langle\mathbf{U}\cdot\mathbf{B}\rangle$, cf.\ e.g.\ Hamba and Tsuchiya (2010), Yokoi (2013); see Yokoi (2023) for a review.

The aim of this paper can be shortly stated as a demonstration of the influence of non-equilibrium effects in MHD turbulence on the $\alpha$-effect and thereby on large-scale dynamos. This issue has already been investigated in a series of papers by Mizerski (2018a,b, 2020, 2021, 2022), which however, assumed that the turbulence was stirred by a Gaussian and helical forcing; the physical properties of the forcing were then present in the expressions for the $\alpha$ coefficient. On the contrary, here we apply the Two-Scale Direct-Interaction Approximation (TSDIA), which allows to remove the stirring force, but instead we need to assume some statistical properties of the background turbulence. Nevertheless, this approach allows to explicitly relate the mean electromotive force to kinetic and cross-helicities, through consideration of the Green response functions, which describe the responses of the turbulent flow and magnetic field to infinitesimal perturbations, cf.\ e.g.\ Yoshizawa (1985, 1990, 1998), Yokoi (2013, 2018). We show that the infinitesimal-impulse cross-responses affect the mean EMF through non-equilibrium effects in MHD turbulence, and the $\alpha$-effect is potentially enhanced, provided that the kinetic and cross-helicities are both non-zero. We also discuss the evolution equation of the cross-helicity, its sources and sinks in MHD turbulence, hence the possibility of a coexistence of the kinetic and cross-helicities; this issue is also investigated numerically.

\section{Mathematical formulation}

To study the magnetohydrodynamic turbulence in an incompressible conducting
fluid we consider the following dynamical equations describing the
evolution of the velocity field of the fluid flow $\mathbf{U}(\mathbf{x},t)$
and the magnetic field $\mathbf{B}(\mathbf{x}.t)$\begin{subequations}
\begin{equation}
\frac{\partial\mathbf{U}}{\partial t}+\left(\mathbf{U}\cdot\nabla\right)\mathbf{U}=-\nabla\Pi-2\boldsymbol{\Omega}\times\mathbf{U}+\left(\mathbf{B}\cdot\nabla\right)\mathbf{B}+\nu\nabla^{2}\mathbf{U},\label{eq:NS}
\end{equation}
\begin{equation}
\frac{\partial\mathbf{B}}{\partial t}+\left(\mathbf{U}\cdot\nabla\right)\mathbf{B}=\left(\mathbf{B}\cdot\nabla\right)\mathbf{U}+\eta\nabla^{2}\mathbf{B},\label{eq:IND}
\end{equation}
\begin{equation}
\nabla\cdot\mathbf{U}=0\qquad\nabla\cdot\mathbf{B}=0,\label{eq:DIVS}
\end{equation}
\end{subequations}where 
\begin{equation}
\Pi=\frac{p}{\rho}+\frac{B^{2}}2-\frac{1}{2}(\boldsymbol{\Omega}\times\mathbf{x})^{2},\label{eq:Pi}
\end{equation}
is the total pressure, $\rho$ is the density,
$\boldsymbol{\Omega}$ is the angular velocity,
$\nu$ is the viscosity, $\eta$ is the magnetic diffusivity.
For the purpose of simplicity we rescaled the magnetic field in the
following way $\mathbf{B}/\sqrt{\mu_{0}\rho}\rightarrow\mathbf{B}$,
where $\mu_0$ is the vacuum permeability
(so that the prefactor $1/\mu_{0}\rho$ in the Lorentz-force term
in the Navier-Stokes equation is lost); in the following we also rescale the currents, $\sqrt{\mu_{0}/\rho}\mathbf{J}\rightarrow\mathbf{J}$, so that $\mathbf{J}=\nabla\times\mathbf{B}$. Next, denoting by angular
brackets the ensemble mean, 
\[
\left\langle \cdot\right\rangle -\textrm{ensemble mean}
\]
we put forward the standard decomposition
\begin{equation}
\mathbf{U}=\left\langle \mathbf{U}\right\rangle +\mathbf{u}',\quad\mathbf{B}=\left\langle \mathbf{B}\right\rangle +\mathbf{b}',\quad p=\left\langle p\right\rangle +p',\label{eq:Reylonds}
\end{equation}
and write down separately the equations for the mean fields $\left\langle \mathbf{U}\right\rangle $
and $\left\langle \mathbf{B}\right\rangle $ and the turbulent fluctuations
$\mathbf{u}'$ and $\mathbf{b'}$; this yields\begin{subequations}
\begin{align}
\frac{\partial\left\langle \mathbf{U}\right\rangle }{\partial t}+\left(\left\langle \mathbf{U}\right\rangle \cdot\nabla\right)\left\langle \mathbf{U}\right\rangle = & -\nabla\left\langle \Pi\right\rangle -2\boldsymbol{\Omega}\times\langle\mathbf{U}\rangle +\left(\left\langle \mathbf{B}\right\rangle \cdot\nabla\right)\left\langle \mathbf{B}\right\rangle +\nu\nabla^{2}\left\langle \mathbf{U}\right\rangle \nonumber \\
 & -\nabla\cdot\left(\left\langle \mathbf{u}'\mathbf{u}'\right\rangle -\left\langle \mathbf{b}'\mathbf{b}'\right\rangle \right),\label{eq:MNS}
\end{align}
\begin{equation}
\frac{\partial\left\langle \mathbf{B}\right\rangle }{\partial t}=\nabla\times\left(\left\langle \mathbf{U}\right\rangle \times\left\langle \mathbf{B}\right\rangle \right)+\nabla\times\left\langle \mathbf{u}'\times\mathbf{b}'\right\rangle +\eta\nabla^{2}\left\langle \mathbf{B}\right\rangle ,\label{eq:MFE}
\end{equation}
\begin{equation}
\nabla\cdot\left\langle \mathbf{B}\right\rangle =0,\quad\nabla\cdot\left\langle \mathbf{U}\right\rangle =0,\label{eq:MDIVS}
\end{equation}
\end{subequations}where 
\begin{equation}
\boldsymbol{\mathcal{E}}=\left\langle \mathbf{u}'\times\mathbf{b}'\right\rangle ,\label{eq:EMF}
\end{equation}
 is the large-scale electromotive force (EMF) and\begin{subequations}
\begin{align}
\frac{\partial\mathbf{u}'}{\partial t}-\nu\nabla^{2}\mathbf{u}'+2\boldsymbol{\Omega}\times\mathbf{u}'+\left(\left\langle \mathbf{U}\right\rangle \cdot\nabla\right)\mathbf{u}'+\left(\mathbf{u}'\cdot\nabla\right)\left\langle \mathbf{U}\right\rangle -\left(\left\langle \mathbf{B}\right\rangle \cdot\nabla\right)\mathbf{b}'-\left(\mathbf{b}'\cdot\nabla\right)\left\langle \mathbf{B}\right\rangle  & \nonumber \\
 & \hspace{-10cm}+\nabla\Pi'=-\nabla\cdot\left(\mathbf{u}'\mathbf{u}'-\mathbf{b}'\mathbf{b}'\right)+\nabla\cdot\left(\left\langle \mathbf{u}'\mathbf{u}'\right\rangle -\left\langle \mathbf{b}'\mathbf{b}'\right\rangle \right),\label{eq:Fluct_u_eq}
\end{align}
\begin{align}
\frac{\partial\mathbf{b}'}{\partial t}-\eta\nabla^{2}\mathbf{b}'+\left(\left\langle \mathbf{U}\right\rangle \cdot\nabla\right)\mathbf{b}'-\left(\left\langle \mathbf{B}\right\rangle \cdot\nabla\right)\mathbf{u}'+\left(\mathbf{u}'\cdot\nabla\right)\left\langle \mathbf{B}\right\rangle -\left(\mathbf{b}'\cdot\nabla\right)\left\langle \mathbf{U}\right\rangle \nonumber\\
&\hspace{-5cm}=\nabla\times\left(\mathbf{u}'\times\mathbf{b}'-\left\langle \mathbf{u}'\times\mathbf{b}'\right\rangle \right),\label{eq:Fluct_b_eq}
\end{align}
\begin{equation}
\nabla\cdot\mathbf{b}'=0,\quad\nabla\cdot\mathbf{u}'=0.\label{eq:fluct_divs}
\end{equation}
\end{subequations}

\section{Non-equilibrium effects in dynamo theory}

Previous results of Mizerski (2018a,b, 2020, 2021, 2022), obtained in the absence of the Coriolis force but with chiral stochastic forcing, in the context
of the geodynamo and galactic dynamos suggest that the non-stationary
$\alpha$-effect is proportional to the energy production rate resulting
from the presence of the forcing (e.g.\ stochastic buoyancy) and is
oscillatory on time scales induced by the forcing, which could be
long (cf.\ also Mizerski \emph{et al}. 2012 for non-stationary dynamo
in the context of the elliptical instability). Here we utilize the
Two-Scale Direct Interaction Approximation, in order to extract the
effect of non-stirred, non-equilibrium turbulence on the large-scale
hydromagnetic dynamo. In other words, the new approach allows to study
non-stationary MHD turbulence and the turbulent dynamo effect in the
absence of external stochastic forcing although with assumed statistical
properties of the background turbulence. We demonstrate, that in non-equilibrium
turbulence the quantity $\left\langle \mathbf{u}'\cdot\mathbf{j}'\right\rangle $
plays a significant role in generation of the large-scale EMF through
the $\alpha$-effect and the effect of $\left\langle \mathbf{u}'\cdot\mathbf{j}'\right\rangle $
vanishes in stationary turbulence.

\subsection{Application of the TSDIA method}

Let us introduce a small parameter $\delta$ and define slow and fast
spatial and temporal variables
\begin{equation}
\boldsymbol{\xi}=\mathbf{x},\quad\mathbf{X}=\delta\mathbf{x},\quad\tau=t,\quad T=\delta t.\label{eq:vars}
\end{equation}
The large-scale fields depend only on the slow variables, $\left\langle \mathbf{U}\right\rangle (\mathbf{X},T)$
and the fluctuations depend on both, $\mathbf{u}^{\prime}(\boldsymbol{\xi},\mathbf{X};\tau,T)$.
We also define the Fourier transform, involving Galilean transformation
to the frame moving with the velocity $\left\langle \mathbf{U}\right\rangle $
\begin{equation}
u_{i}^{\prime}(\boldsymbol{\xi},\mathbf{X};\tau,T)=\int\mathrm{d}^{3}k\hat{u}_{i}^{\prime}(\mathbf{k},\mathbf{X;\tau},T)\mathrm{e}^{-\mathrm{i}\mathbf{k}\cdot(\boldsymbol{\xi}-\left\langle \mathbf{U}\right\rangle \tau)},\label{eq:F_transform_u}
\end{equation}
but the explicit dependence on the slow variables $\mathbf{X}$ and
$T$ will be typically suppressed in notation for clarity. The details of the TSDIA approach are provided in Appendix~A (see also $\mathsection$ 9.6 of Yoshizawa 1998, Yoshizawa 1985, 1990 and Yokoi 2023) and here we present the major results.
The method involves introduction of the concept of background turbulence with given statistical properties, uninfluenced by the
large-scale field and rotation, hence isotropic; this background turbulence is defined
by the following correlation functions
\begin{equation}
\left\langle \hat{f}_{i}(\mathbf{k};\tau)\hat{g}_{j}(\mathbf{k}_{1};\tau_{1})\right\rangle =\left[P_{ij}(\mathbf{k})Q_{fg}\left(k;\tau,\tau_{1}\right)+\frac{1}{2}\mathrm{i}\epsilon_{ijk}\frac{k_{k}}{k^{2}}H_{fg}\left(k;\tau,\tau_{1}\right)\right]\delta(\mathbf{k}+\mathbf{k}_{1}),\label{eq:correlations}
\end{equation}
\begin{equation}
\left\langle G_{fgij}^{\prime}(\mathbf{k};\tau,\tau_{1})\right\rangle =\delta_{ij}G_{fg}\left(k;\tau,\tau_{1}\right),\label{eq:correlations_G}
\end{equation}
where $f$ and $g$ represent one of the variables $\mathbf{u}_{00}^{\prime}$
and $\mathbf{b}_{00}^{\prime}$ and $G_{fij}^{\prime}(\mathbf{k};\tau,\tau_{1})$ denote the Green's functions describing the system's response to infinitesimal disturbances. It is useful at this stage to write down explicitly the following quantity
\begin{align}
\left\langle \mathbf{u}_{00}^{\prime}(\mathbf{x},\tau)\cdot\mathbf{j}_{00}^{\prime}(\mathbf{x},\tau_{1})\right\rangle &\,\,= -\mathrm{i}\epsilon_{ijk}\int\mathrm{d}k\int\mathrm{d}k^{\prime}k^{\prime}_j\left\langle \hat{u}_{00i}^{\prime}(\mathbf{k};\tau)\hat{b}_{00k}^{\prime}(\mathbf{k}^{\prime};\tau_{1})\right\rangle\mathrm{e}^{-\mathrm{i}(\mathbf{k}+\mathbf{k}^{\prime})\cdot\mathbf{x}} \nonumber\\
&\,\,= \int\mathrm{d}k H_{ub}\left(k;\tau,\tau_{1}\right)=\int\mathrm{d}k H_{bu}\left(k;\tau_1,\tau\right),\label{eq:correlations-2}
\end{align}
since this quantity will play an important role in the theory of non-equilibrium $\alpha$-effect, developed below.

The derivation of the formula for the EMF presented in Appendix~A leads to
\begin{equation}
\boldsymbol{\mathcal{E}}=\alpha\left\langle \mathbf{B}\right\rangle -\left(\beta+\zeta\right)\left\langle\mathbf{J}\right\rangle 
-\nabla\zeta\times\left\langle \mathbf{B}\right\rangle
+\gamma\left(\left\langle \mathbf{W}\right\rangle +2\boldsymbol{\Omega}\right),\label{eq:EMF_gen}
\end{equation}
where $\mathbf{J}=\nabla\times\mathbf{B}=\left\langle \mathbf{J}\right\rangle +\mathbf{j}$'
and $\mathbf{W}=\nabla\times\mathbf{U}=\left\langle \mathbf{W}\right\rangle +\mathbf{w}'$
denote electric currents and the vorticity respectively. The statistically
stationary case has been studied in detail in Yoshizawa (1998) and
Yokoi (2013, 2018). 

We now concentrate on the $\alpha$-effect, which can be decomposed into two contributions, 
\begin{equation}
\alpha=\alpha_{S}+\alpha_{\rm X},\label{eq:alpha}
\end{equation}
the standard one, related to the so-called residual helicity
\begin{align}
\alpha_{S}=& \frac{1}{3}\int\mathrm{d}^{3}k\int_{-\infty}^{\tau}\mathrm{d}\tau_{1}\left[G_{uu}\left(k,\mathbf{X};\tau,\tau_{1},T\right)H_{bb}\left(k,\mathbf{X};\tau,\tau_{1},T\right)\right.\nonumber\\
&\left.\qquad\qquad\qquad\qquad-G_{bb}\left(k,\mathbf{X};\tau,\tau_{1},T\right)H_{uu}\left(k,\mathbf{X};\tau_{1},\tau,T\right)\right],\label{eq:alpha_s}
\end{align}
and a less obvious one, related to the cross helicity and the quantity $\langle\mathbf{u}'\cdot\mathbf{j}'\rangle$ which takes the form
\begin{align}
\alpha_{\rm X}= & -\frac{1}{3}\int\mathrm{d}^{3}k\int_{-\infty}^{\tau}\mathrm{d}\tau_{1}G_{bu}\left(k,\mathbf{X};\tau,\tau_{1},T\right)H_{ub}\left(k,\mathbf{X};\tau,\tau_{1},T\right)\nonumber \\
 & +\frac{1}{3}\int\mathrm{d}^{3}k\int_{-\infty}^{\tau}\mathrm{d}\tau_{1}G_{ub}\left(k,\mathbf{X};\tau,\tau_{1},T\right)H_{bu}\left(k,\mathbf{X};\tau,\tau_{1},T\right).\label{eq:alpha_tilde}
\end{align}
Since the helical functions of the background turbulence satisfy
\begin{equation}
H_{bu}\left(\tau,\tau_{1}\right)=H_{ub}\left(\tau_{1},\tau\right),\label{eq:proprty_of_H}
\end{equation}
we obtain
\begin{align}
\alpha_{\rm X}= & -\frac{1}{3}\int\mathrm{d}^{3}k\int_{-\infty}^{\tau}\mathrm{d}\tau_{1}G_{bu}\left(k,\mathbf{X};\tau,\tau_{1},T\right)H_{ub}\left(k,\mathbf{X};\tau,\tau_{1},T\right)\nonumber \\
 & +\frac{1}{3}\int\mathrm{d}^{3}k\int_{-\infty}^{\tau}\mathrm{d}\tau_{1}G_{ub}\left(k,\mathbf{X};\tau,\tau_{1},T\right)H_{ub}\left(k,\mathbf{X};\tau_{1},\tau,T\right).\label{eq:alpha_tilde_1}
\end{align}
We now introduce the following symmetric and antisymmetric parts of $H_{ub}$ with respect to exchange of time variables
\begin{subequations}
\begin{equation}
H_{ub}^{(s)}\left(\tau,\tau_{1}\right) = \frac{1}{2}\left( H_{ub}\left(\tau,\tau_{1}\right) + H_{ub}\left(\tau_1,\tau\right) \right),
\label{sym}
\end{equation}
\begin{equation}
H_{ub}^{(a)}\left(\tau,\tau_{1}\right) = \frac{1}{2}\left( H_{ub}\left(\tau,\tau_{1}\right) - H_{ub}\left(\tau_1,\tau\right) \right),
\label{antisym}
\end{equation}
\end{subequations}
which allows to further separate the $\alpha_{\rm X}$ term into two contributions
\begin{align}
\alpha_{\rm X}= & -\frac{1}{3}\int\mathrm{d}^{3}k\int_{-\infty}^{\tau}\mathrm{d}\tau_{1}\left[G_{ub}\left(k,\mathbf{X};\tau,\tau_{1},T\right) + G_{bu}\left(k,\mathbf{X};\tau,\tau_{1},T\right)\right]H_{ub}^{(a)}\left(k,\mathbf{X};\tau,\tau_{1},T\right)\nonumber \\
 & +\frac{1}{3}\int\mathrm{d}^{3}k\int_{-\infty}^{\tau}\mathrm{d}\tau_{1}\left[G_{ub}\left(k,\mathbf{X};\tau,\tau_{1},T\right) - G_{bu}\left(k,\mathbf{X};\tau,\tau_{1},T\right)\right]H_{ub}^{(s)}\left(k,\mathbf{X};\tau_{1},\tau,T\right).\label{eq:alpha_tilde_sa}
\end{align}
The first term in equation (\ref{eq:alpha_tilde_sa}), i.e.\
\begin{equation}
\alpha_{\rm neq}=  -\frac{1}{3}\int\mathrm{d}^{3}k\int_{-\infty}^{\tau}\mathrm{d}\tau_{1}\left[G_{ub}\left(k,\mathbf{X};\tau,\tau_{1},T\right) + G_{bu}\left(k,\mathbf{X};\tau,\tau_{1},T\right)\right]H_{ub}^{(a)}\left(k,\mathbf{X};\tau,\tau_{1},T\right),\label{eq:alpha_neq}
\end{equation}
clearly constitutes a contribution from non-stationarity of
the turbulence, as the antisymmetric part $H_{ub}^{(a)}$ is clearly a non-equilibrium effect. 

\subsection{Physics of the non-equilibrium $\alpha_{\rm neq}$-effect}

If we further assume that the function
\begin{equation}
\mathcal{G}\left(\tau,\tau_1\right) = G_{ub}\left(\tau,\tau_{1}\right) + G_{bu}\left(\tau,\tau_{1}\right)
\end{equation}
is independent of $k$, the non-equilibrium $\alpha$-effect can be expressed as follows:
\begin{equation}
\alpha_{\rm neq}=  -\frac{1}{3}\int_{-\infty}^{\tau}\mathrm{d}\tau_{1}\mathcal{G}\left(\tau,\tau_1\right)\langle\mathbf{u}_{00}^{\prime}\cdot\mathbf{j}_{00}^{\prime}\rangle^{(a)}\left(\mathbf{x},\tau,\tau_1\right),\label{eq:alpha_neq_2}
\end{equation}
where
\begin{equation}
\langle\mathbf{u}_{00}^{\prime}\cdot\mathbf{j}_{00}^{\prime}\rangle^{(a)}\left(\mathbf{x},\tau,\tau_1\right) = \frac{1}{2}\left[\langle\mathbf{u}_{00}^{\prime}\left(\mathbf{x},\tau\right)\cdot\mathbf{j}_{00}^{\prime}\left(\mathbf{x},\tau_1\right)\rangle - \langle\mathbf{u}_{00}^{\prime}\left(\mathbf{x},\tau_1\right)\cdot\mathbf{j}_{00}^{\prime}\left(\mathbf{x},\tau\right)\rangle\right].\label{eq:antisym_uj}
\end{equation}
The memory effect, expressed by the time integral in (\ref{eq:antisym_uj}) is clearly crucial, as $\langle\mathbf{u}_{00}^{\prime}\cdot\mathbf{j}_{00}^{\prime}\rangle^{(a)}\left(\mathbf{x},\tau,\tau\right)=0$. Next, inspection of the evolution
equations for the Green's functions leads to the conclusion that $G_{ub}$
must be an odd function of $\mathbf{b}_{00}^{\prime}$. This is expected,
since the $\alpha_{\rm X}$ contribution to the $\alpha$-effect results
from the action of the Lorentz force, and since $H_{ub}$ is associated
with the quantity $\left\langle \mathbf{u}_{00}^{\prime}\cdot\mathbf{j}_{00}^{\prime}\right\rangle $, i.e.\ $H_{ub}$
is linear in $\mathbf{b}_{00}^{\prime}$, it follows that $G_{ub}$
must be an odd function of the latter. Moreover, since $\left\langle \mathbf{u}_{00}^{\prime}\cdot\mathbf{j}_{00}^{\prime}\right\rangle $
is a scalar quantity (does not change sign under reflections), $G_{ub}$
must be skew. The only dynamical quantity that is skew and odd in $\mathbf{b}_{00}^{\prime}$
is the cross helicity, $\langle\mathbf{u}_{00}^{\prime}\cdot\mathbf{b}_{00}^{\prime}\rangle$, hence we expect that $G_{ub}\sim Q_{ub}$. Having in mind that the response function $\mathcal{G}(\tau,\tau_1)$ is non-dimensional we can now provide the following rough estimate of the non-equilibrium $\alpha_{\rm neq}$-effect
\begin{equation}
\alpha_{\rm neq}\sim  -\frac{2}{3}\int_{-\infty}^{\tau}\mathrm{d}\tau_{1}\Upsilon^{(s)}\left(\mathbf{x},\tau,\tau_1\right)\langle\mathbf{u}_{00}^{\prime}\cdot\mathbf{j}_{00}^{\prime}\rangle^{(a)}\left(\mathbf{x},\tau,\tau_1\right),\label{eq:alpha_neq_est}
\end{equation}
where
\begin{equation}
\Upsilon\left(\mathbf{x},\tau,\tau_1\right) =  \frac{\langle\mathbf{u}_{00}^{\prime}\left(\mathbf{x},\tau\right)\cdot\mathbf{b}_{00}^{\prime}\left(\mathbf{x},\tau_1\right)\rangle}{\sqrt{\langle u_{00}^{\prime 2}\rangle\left(\mathbf{x},\tau\right)\langle b_{00}^{\prime 2}\rangle\left(\mathbf{x},\tau_1\right)}},\label{eq:nondim_CH}
\end{equation}
\begin{equation}
\Upsilon^{(s)}\left(\mathbf{x},\tau,\tau_1\right) = \frac{1}{2}\left[ \Upsilon\left(\mathbf{x},\tau,\tau_1\right) + \Upsilon\left(\mathbf{x},\tau_1,\tau\right) \right] ,\label{eq:CH_sym}
\end{equation}
and the cross helicity has been normalized by the geometric mean of the kinetic and magnetic fluctuational energies (see Yokoi 2011 for a discussion of different cross-helicity normalizations). The latter equation expresses an effect which results from the lack of
equilibrium in the turbulent state.

The second term in (\ref{eq:alpha_tilde_sa}) is likely to be small because of the factor $G_{ub}(\tau,\tau_1)-G_{bu}(\tau,\tau_1)$. For example in the case when $\nu=\eta$ the two response functions $G_{ub}$ and $G_{bu}$ are equal and $\alpha_{\rm X}=\alpha_{\rm neq}$. This still holds approximately true, when the diffusivities are unequal but weak, 
\begin{equation}
G_{ub}\approx G_{bu},\qquad\textrm{and}\qquad\alpha_{\rm X}\approx\alpha_{\rm neq}.\label{eq:GubGbu_aneq}
\end{equation}
The same symmetry arguments as in the case of $\alpha_{\rm neq}$ can be also applied to the second term in (\ref{eq:alpha_tilde_sa}) which is therefore proportional to the non-dimensional cross helicity $\Upsilon=\Upsilon(\mathbf{x},\tau,\tau)$ and the quantity $\langle\mathbf{u}_{00}^{\prime}\cdot\mathbf{j}_{00}^{\prime}\rangle=\langle\mathbf{u}_{00}^{\prime}\left(\mathbf{x},\tau\right)\cdot\mathbf{j}_{00}^{\prime}\left(\mathbf{x},\tau\right)\rangle$, i.e.\
$\alpha_{\rm X}-\alpha_{\rm neq}\sim \tau_{t}\Upsilon\left\langle \mathbf{u}_{00}^{\prime}\cdot\mathbf{j}_{00}^{\prime}\right\rangle$, where $\tau_t$ is the turn over time of the most energetic turbulent eddies. However, as remarked above this effect should be weak, when the diffusion is weak or the magnetic Prandtl number $\mbox{Pr}_M=\nu/\eta\approx1$.

Finally, we also expect the $\langle\mathbf{u}^{\prime}_{00}\cdot\mathbf{j}^{\prime}_{00}\rangle$ correlations in fully turbulent flows to be proportional to the kinetic helicity $\langle\mathbf{u}^{\prime}_{00}\cdot\mathbf{w}^{\prime}_{00}\rangle$, since typically the velocities and magnetic fields tend to align in such flows. Again, the prefactor must be skew and odd in $\mathbf{b}^{\prime}_{00}$, therefore we propose 
\begin{equation}
\langle\mathbf{u}^{\prime}_{00}\cdot\mathbf{j}^{\prime}_{00}\rangle \approx \Upsilon \langle\mathbf{u}^{\prime}_{00}\cdot\mathbf{w}^{\prime}_{00}\rangle = \frac{\langle\mathbf{u}^{\prime}_{00}\cdot\mathbf{b}^{\prime}_{00}\rangle
\langle\mathbf{u}^{\prime}_{00}\cdot\mathbf{w}^{\prime}_{00}\rangle}{\sqrt{\left\langle u^{\prime2}_{00}\right\rangle \left\langle b^{\prime2}_{00}\right\rangle} }.
\label{uj_through_uo}
\end{equation}
Introducing the latter relation into (\ref{eq:alpha_neq_est}) leads to
\begin{equation}
\alpha_{\rm neq}\sim  -\frac{2}{3}\int_{-\infty}^{\tau}\mathrm{d}\tau_{1}\left(\Upsilon^{(s)}\left(\mathbf{x},\tau,\tau_1\right)\right)^2\langle\mathbf{u}_{00}^{\prime}\cdot\mathbf{w}_{00}^{\prime}\rangle^{(a)}\left(\mathbf{x},\tau,\tau_1\right),\label{eq:alpha_neq_est_fin}
\end{equation}
which shows, that the non-equilibrium $\alpha_{\rm neq}$-effect relies on coexistence of the kinetic and cross helicities and their history in MHD turbulence (more precisely, in the case of kinetic helicity only the antisymmetric part of the time correlations $\langle\mathbf{u}_{00}^{\prime}\cdot\mathbf{w}_{00}^{\prime}\rangle^{(a)}\left(\mathbf{x},\tau,\tau_1\right)$ contributes to the new effect). 

\subsection{Calculation of the $\alpha_{\rm neq}$-effect}

We will now investigate this dynamo mechanism in some more detail. In order to calculate the effect of non-equilibrium
turbulence we adopt a similar approach to that in $\mathsection$
7 of Yoshizawa (1998). In stationary turbulence the functions $H_{fg}\left(k,\mathbf{X};\tau,\tau_{1},T\right)$
and $G_{f}\left(k,\mathbf{X};\tau,\tau_{1},T\right)$ depend only
on $\left|\tau-\tau_{1}\right|$, hence to study the non-equilibrium
effects we postulate a similar formulae for these functions as those
of Yoshizawa (1998) (cf.\ formulae 6.53-6.54 of this book), but modified
in order to introduce simple explicit and distinct dependencies on
$\tau$ and $\tau_{1}$
\begin{equation}
H_{fg}\left(k,\mathbf{k}\cdot\left\langle \mathbf{B}\right\rangle ,\mathbf{X};\tau,\tau_{1},T\right)=\sigma\left(k,\mathbf{X},T\right)\mathrm{e}^{-\varpi\left(k,\mathbf{X},T\right)\left|\tau-\tau_{1}\right|}\mathcal{H}\left(\tau\right)\mathcal{H}_{1}\left(\tau_{1}\right),\label{eq:Hub_gen}
\end{equation}
\begin{equation}
G_{fg}\left(k,\mathbf{X};\tau,\tau_{1},T\right)=\theta\left(\tau-\tau_{1}\right)\varsigma\left(k,\mathbf{X},T\right)\mathrm{e}^{-\varpi\left(k,\mathbf{X},T\right)\left|\tau-\tau_{1}\right|}\mathcal{G}\left(\tau\right)\mathcal{G}_{1}\left(\tau_{1}\right),\label{eq:GA_gen}
\end{equation}
for some functions $\mathcal{H}\left(\tau\right)$, $\mathcal{H}_{1}(\tau_{1})$,
$\mathcal{G}(\tau)$ and $\mathcal{G}_{1}(\tau_{1})$. We can decompose these functions
into Fourier modes, which allows to adopt the following, simpler, generic model\begin{subequations}
\begin{equation}
H_{fg}\left(\tau,\tau_{1}\right)=\sigma\mathrm{e}^{-\varpi\left|\tau-\tau_{1}\right|}\sin\left(\varpi_{h0}\tau\right)\sin\left(\varpi_{h1}\tau_{1}\right),\label{eq:Hub_onemode}
\end{equation}
\begin{equation}
G_{fg}\left(\tau,\tau_{1}\right)=\theta\left(\tau-\tau_{1}\right)\varsigma\mathrm{e}^{-\varpi\left|\tau-\tau_{1}\right|}\sin\left(\varpi_{g0}\tau\right)\sin\left(\varpi_{g1}\tau_{1}\right),\label{eq:GA_onemode}
\end{equation}
\end{subequations}where the dependence on the slow variables and
the wavenumber $k$ was suppressed in notation for clarity; moreover
$\varpi>0$ and to fix ideas we also assume $\varpi_{h0}>0$, $\varpi_{g0}>0$,
$\varpi_{h1}>0$ and $\varpi_{g1}>0$. For the sake of simplicity we also assume 
\begin{equation}
G_{ub}\approx G_{bu}.\label{eq:GubGbu_2}
\end{equation}
The following calculation
\begin{align}
\int_{-\infty}^{\tau}\mathrm{d}\tau_{1}G_{fg}\left(\tau,\tau_{1}\right)H_{fg}\left(\tau,\tau_{1}\right)\nonumber \\
 & \hspace{-3cm}=\frac{\sigma\varsigma}{4}\left(\cos\Delta_{0}\tau-\cos\Sigma_{0}\tau\right)\left[\frac{1}{4\varpi^{2}+\Delta_{1}^{2}}\left(2\varpi\cos\Delta_{1}\tau+\Delta_{1}\sin\Delta_{1}\tau\right)\right.\qquad\qquad\qquad\qquad\qquad\qquad\qquad\nonumber\\
 &\hspace{-3cm}\qquad\qquad\qquad\qquad\qquad\qquad\left.-\frac{1}{4\varpi^{2}+\Sigma_{1}^{2}}\left(2\varpi\cos\Sigma_{1}\tau+\Sigma_{1}\sin\Sigma_{1}\tau\right)\right],\label{eq:GH_est}
\end{align}
where
\begin{equation}
\Delta_{i}=\varpi_{hi}-\varpi_{gi},\qquad\Sigma_{i}=\varpi_{hi}+\varpi_{gi},\label{eq:DeltaSigma}
\end{equation}
shows, that in non-equilibrium turbulence both contributions to the
$\alpha$-effect, the `standard' $\alpha_{S}$ and the one associated
with cross helicity $\alpha_{X}$, are enhanced by non-stationarity.
Since the frequencies correspond to the fast oscillations of turbulent
fluctuations in most of the cases the cosines and sines do not
contribute to large time scales (their time average vanishes). Under
the time average over long time scales $\delta^{-1}t$ the non-zero
contribution comes from the cases $\varpi_{hi}=\varpi_{gi}$ (or $\varpi_{hi}\approx\varpi_{gi}$).
Therefore we pick $(\varpi,\,\varpi_{h},\,\varpi_{g})$-modes such
that the following relations are satisfied
\begin{equation}
\Delta_{i}\ll\varpi\ll\varpi_{hi},\,\varpi_{gi},\quad\textrm{for}\quad i=0,1,\label{eq:simpl_assumpt}
\end{equation}
in which case 
\begin{equation}
\int_{-\infty}^{\tau}\mathrm{d}\tau_{1}G_{fg}\left(\tau,\tau_{1}\right)H_{fg}\left(\tau,\tau_{1}\right)\approx\frac{\sigma\varsigma}{8\varpi};\label{eq:non-stat_gen}
\end{equation}
for comparison in the stationary case one obtains $\sigma_{s}\varsigma_{s}/2\varpi_{s}$
with $H_{fg}=\sigma_{s}\exp(-\varpi_{s}\left|\tau-\tau_{1}\right|)$,
$G_{fg}=\varsigma_{s}\exp(-\varpi_{s}\left|\tau-\tau_{1}\right|)$.
However, the influence of non-stationarity on the `standard' $\alpha_{S}$
contribution has been studied using different methods in Mizerski
(2018a,b, 2020, 2021, 2022). Here we concentrate on the cross-helicity
contribution $\alpha_{\rm X}\approx\alpha_{\rm neq}$, which is apparent within the TSDIA
approach. Introduction of the formulae (\ref{eq:Hub_onemode},b) into
(\ref{eq:alpha_neq}) yields
\begin{equation}
\alpha_{\rm neq}\approx-\frac{\pi}{6}\int\mathrm{d}k\frac{\sigma\varsigma k^{2}}{\varpi}.\label{eq:alpha_ns_1}
\end{equation}
According to our previous observations in the above we have $\varsigma\sim \Upsilon$.
We note that a very similar result is obtained if one assumes a simpler
non-stationary form of the $H_{ub}$ and $G_{ub}$ functions\begin{subequations}
\begin{equation}
H_{ub}\left(\tau,\tau_{1}\right)=\sigma\mathrm{e}^{-\varpi\left|\tau-\tau_{1}\right|}\sin\left[\varpi_{h}\left(\tau-\tau_{1}\right)\right],\label{eq:Hub_onemode-1}
\end{equation}
\begin{equation}
G_{ub}\left(\tau,\tau_{1}\right)=\theta\left(\tau-\tau_{1}\right)\varsigma\mathrm{e}^{-\varpi\left|\tau-\tau_{1}\right|}\sin\left[\varpi_{g}\left(\tau-\tau_{1}\right)\right],\label{eq:GA_onemode-1}
\end{equation}
\end{subequations}which satisfies $H_{ub}(\tau,\tau_1)=-H_{ub}(\tau_1,\tau)$, and considers the limit (\ref{eq:simpl_assumpt}).

In the above calculation we have used some standard models of the statistical properties of turbulence in order to emphasize the importance of the history of evolution of the helicities in the turbulent dynamo process.  The $\alpha_{\rm neq}$-effect, induced by the simultaneous presence of cross and kinetic helicities, can be strong and depends on their magnitude.

\section{Coexistence of the kinetic and cross helicities in turbulence}\label{subsec:coexist}

We now consider the question of the likelihood of coexistence of the cross and kinetic helicities in developed turbulence. Although it is not possible to draw definite conclusions in this matter, it is still instructive to study the sources and sinks of the cross helicity in turbulent flows in order to develop some intuition about its generation.

In the Appendix~B we consider a stirred turbulence (with homogeneous, isotropic, stationary and helical Gaussian forcing) and show that under the first-order smoothing approximation the kinetic helicity is proportional to the helicity of the forcing, whereas the  cross-helicity is defined by the product $\langle\mathbf{f}\cdot\nabla\times\mathbf{f}\rangle(\langle\mathbf{B}\rangle\cdot\boldsymbol{\Omega})$. In other words, within the FOSA approach the existence of the cross-helicity is dependent on the existence of the mean field component parallel to the background rotation vector. 

A more general calculation is presented in the Appendix~C, where we have derived the general evolution equation for the cross-helicity (cf.\ also Yokoi and Hamba 2007, Yokoi 2011, Yokoi and Balarac 2011, Yokoi and Hoshino 2011, Yokoi 2013). This equation involves mean quantities such as the mean EMF $\boldsymbol{\mathcal{E}}$ and the mean Reynolds and Maxwell stresses $\langle u_i^{\prime} u_j^{\prime} -  b_i^{\prime} b_j^{\prime}\rangle$. For the former we utilize the result (\ref{eq:EMF_gen}) and for the latter we take the expression obtained also via the TSDIA approach in Yokoi and Hoshino (2011), i.e.\
\begin{equation}
-\langle u_i^{\prime} u_j^{\prime} -  b_i^{\prime} b_j^{\prime}\rangle\frac{\partial\langle B\rangle_i}{\partial x_j} = \frac{7}{10}\beta \mathcal{S}_{ij}\mathcal{M}_{ij} - \frac{7}{10}\gamma \mathrm{Tr}\left(\boldsymbol{\mathcal{M}}^2\right),\label{ReyMax}
\end{equation}
where 
\begin{equation}
\mathcal{S}_{ij} = \frac{\partial \langle U\rangle_i}{\partial x_j} + \frac{\partial \langle U\rangle_j}{\partial x_i},\qquad \mathcal{M}_{ij} = \frac{\partial \langle B\rangle_i}{\partial x_j} + \frac{\partial \langle B\rangle_j}{\partial x_i}.\label{SM}
\end{equation}
This leads to
\begin{align}
\frac{D}{Dt}\left\langle \mathbf{u}^{\prime}\cdot\mathbf{b}^{\prime}\right\rangle = & -\alpha\left(\left\langle \mathbf{B}\right\rangle \cdot\left\langle \mathbf{W}\right\rangle +2\left\langle \mathbf{B}\right\rangle \cdot\boldsymbol{\Omega}\right)+\left(\beta+\zeta\right) \left(\left\langle \mathbf{J}\right\rangle \cdot\left\langle \mathbf{W}\right\rangle +2\left\langle \mathbf{J}\right\rangle \cdot\boldsymbol{\Omega}\right)\nonumber\\
 &\hspace{-1.5cm} -\gamma\left(\left\langle \mathbf{W}\right\rangle +2\boldsymbol{\Omega}\right)^{2}-\frac{7}{10}\gamma \mathrm{Tr}\left(\boldsymbol{\mathcal{M}}^2\right)\nonumber\\
 & \hspace{-1.5cm}+\frac{7}{10}\beta \mathrm{Tr}\left(\boldsymbol{\mathcal{S}}\cdot\boldsymbol{\mathcal{M}}\right)+\left(\nabla\zeta\times\left\langle \mathbf{B}\right\rangle\right)\cdot\left(\left\langle \mathbf{W}\right\rangle +2\boldsymbol{\Omega}\right)\nonumber\\
 & \hspace{-1.5cm}+\nabla\cdot\left[\left\langle \left(-\Pi'+\frac{\mathbf{u}^{\prime2}+\mathbf{b}^{\prime2}}{2}\right)\mathbf{b}'\right\rangle +\left\langle \frac{\mathbf{u}^{\prime2}+\mathbf{b}^{\prime2}}{2}\right\rangle \left\langle \mathbf{B}\right\rangle -\nu\left\langle \mathbf{w}'\times\mathbf{b}'\right\rangle +\eta\left\langle \mathbf{u}'\times\mathbf{j}'\right\rangle \right]\nonumber\\
 & \hspace{-1.5cm}-\left(\nu+\eta\right)\left\langle \mathbf{w}'\cdot\mathbf{j}^{\prime}\right\rangle. \label{Hprod}
\end{align}
Of course if the turbulence is stirred with some forcing $\mathbf{f}$ there is  also another production  term $\left\langle \mathbf{f}\cdot\mathbf{b}'\right\rangle$. 

According to
(\ref{eq:alpha_neq_est_fin}) the magnitude of the non-equilibrium $\alpha$-effect depends on both, the kinetic and cross helicities and their history.
The total $\alpha$-effect consists of the two contributions $\alpha=\alpha_{\rm S}+\alpha_{\rm X}$, where the standard one can be assumed proportional to the kinetic helicity, $\alpha_{\rm S}\approx-\tau_t \langle \mathbf{u}'\cdot\mathbf{w}' \rangle/3$. The final balance between the two contributions $\alpha_{\rm S}$ and $\alpha_{\rm X}$ determines whether the $\alpha$ coefficient has the same or the opposite sign to the kinetic helicity. The effect of different terms in the equation (\ref{Hprod}) has been studied in the aforementioned works of Yokoi and Hamba (2007), Yokoi (2011), Yokoi and Balarac (2011), Yokoi and Hoshino (2011) and Yokoi (2013) under some simplifying assumptions, in particular under the neglect of the effects from the $G_{ub}$ and $G_{bu}$ response functions, responsible for the non-equilibrium effects studied here. Assuming that $\alpha=-\tau_t\langle \mathbf{u}'\cdot\mathbf{w}' \rangle/3$ they showed, that the first term $-\alpha\langle\mathbf{B}\rangle\cdot\langle\mathbf{W}\rangle$ always leads to destruction of the cross helicity. This is no longer true, when $\Upsilon\neq0$ in non-equilibrium turbulence, since depending on the balance between the $\alpha_{\rm S}$ and $\alpha_{\rm X}$ terms the term $-\alpha\langle\mathbf{B}\rangle\cdot\langle\mathbf{W}\rangle$ in (\ref{Hprod}) may either amplify or destroy the cross helicity. Furthermore, Yokoi and Hoshino (2011) take $\beta+\zeta \sim \langle \mathbf{u}^{\prime 2} \rangle$ and $\gamma \sim \langle \mathbf{u}'\cdot\mathbf{b}' \rangle$  
which allows them to identify another two terms that  always lead to destruction of the cross-helicity, namely %
\begin{equation}
-\gamma\left(\left\langle \mathbf{W}\right\rangle +2\boldsymbol{\Omega}\right)^{2}-\frac{7}{10}\gamma \mathrm{Tr}\left(\boldsymbol{\mathcal{M}}^2\right). \label{destruct}
\end{equation}
In addition Yokoi and Hoshino (2011) have described various situations when the terms $\left(\beta+\zeta\right) \left\langle \mathbf{J}\right\rangle \cdot\left\langle \mathbf{W}\right\rangle$, $\beta \mathrm{Tr}\left(\boldsymbol{\mathcal{S}}\cdot\boldsymbol{\mathcal{M}}\right)$ and $\nabla\cdot\left[\left\langle \mathbf{u}^{\prime2}+\mathbf{b}^{\prime2}\right\rangle \left\langle \mathbf{B}\right\rangle \right]$ may lead to production of the cross-helicity in the geometry of the tokamak devices. Finally, in the term $-2\alpha\left\langle \mathbf{B}\right\rangle \cdot\boldsymbol{\Omega}$ we recover the action of the mean field component parallel to the rotation vector, as in the FOSA approach.

The action of all the other terms in (\ref{Hprod}) is difficult to predict and, in general, they can either amplify or destroy the cross-helicity in developed turbulence. The final balance on the right hand side of (\ref{Hprod}) depends on many dynamical features of turbulence and is expected to be time dependent. Therefore in order to demonstrate the possibility of coexistence of the cross- and kinetic helicities in magnetized turbulence we have performed numerical simulations of the compressible version of equations (\ref{eq:NS}--c) in the presence of gravity, density stratification and an imposed magnetic field $\mathbf{g}\parallel\nabla\rho\parallel\mathbf{B}_0\parallel\boldsymbol{\Omega}$ in a periodic box with the use of the {\sc Pencil Code} (Pencil Code Collaboration) with $256^3$ mesh points; stress-free and perfectly conducting boundary conditions were imposed at the top and bottom boundaries; see Appendix~D.
The action of rotation along the direction of stratification leads to kinetic helicity (see figure~5 of Jabbari et al.\ 2014 for simulation results)
and the action of a magnetic field along the direction of stratification leads to cross helicity (R\"udiger et al.\ 2011).

\begin{figure}
\begin{centering}
\includegraphics[width=0.7\linewidth]{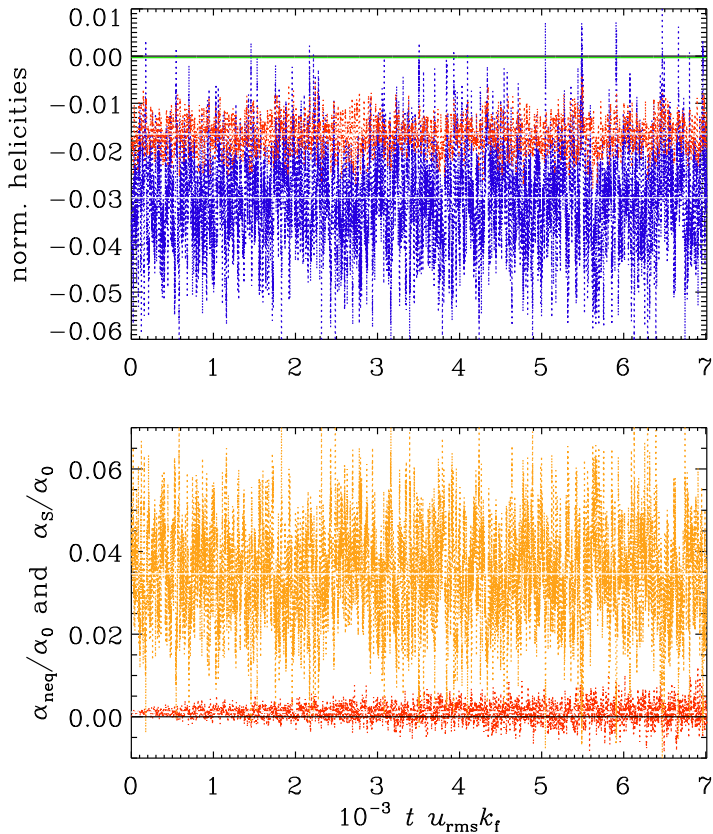}
\par\end{centering}
\caption{\label{fig:b0-01} Results for Run~A of numerical simulations of MHD turbulence in a periodic box with the use of the {\sc{Pencil Code}}. The upper panel shows the time evolution of the normalized cross-helicity, $\langle\mathbf{u}^{\prime}\cdot\mathbf{b}^{\prime}\rangle/\sqrt{\langle u^{\prime 2}\rangle\langle b^{\prime 2}\rangle}$ (red) and the kinetic helicity $\langle\mathbf{u}^{\prime}\cdot\mathbf{w}^{\prime}\rangle/\sqrt{\langle u^{\prime 2}\rangle\langle w^{\prime 2}\rangle}$ (blue); the time averages are marked with the white continuous lines and the green line depicts the current helicity $\langle\mathbf{b}^{\prime}\cdot\mathbf{j}^{\prime}\rangle/\sqrt{\langle u^{\prime 2}\rangle\langle w^{\prime 2}\rangle}$. The estimates of the coefficients $\alpha_{\rm neq}$ (\ref{est_alpha_neq}) and $\alpha_{\rm S}$ (\ref{est_alpha_standard}) as functions of time (normalized with $\alpha_0 = u_{\rm rms}/3$) are provided in the bottom panel in red and orange respectively; the continuous white line marks the time averaged value of $\alpha_{\rm S}/\alpha_0$ and the dashed white line the time average of $\alpha_{\rm neq}/\alpha_0$.
}
\end{figure}

The values of the physical parameters are as follows:
working again with the unscaled magnetic field
$B=0.01\,c_{\rm s}\sqrt{\mu_0\bar{\rho}}$ and gravity $g=1\, c_{\rm s}^2 k_1$
(these are varied in other runs), where $c_{\rm s}$ is the speed of sound, $\Omega=0.5\,c_{\rm s} k_1$ is kept fixed in all runs, $\bar{\rho}$ is the mean density and $k_1$ is the box wavenumber; the remaining parameters, which are constant for all runs are listed in table~1, where we used the Alfv\'en speed $v_A=B/\sqrt{\mu_0\overline{\rho}}$ to quantify the strength of the imposed and rms magnetic fields through $v_{A0}$ and $v_{A}^{\rm rms}$, respectively.
The results obtained for two values of the imposed magnetic field which
differ by an order of magnitude at variable gravity strength are depicted
in figure \ref{fig:b0-01} and tables~1 and~2; see also Appendix~E for additional figures.
The normalized helicities, $\langle\mathbf{u}^{\prime}\cdot\mathbf{b}^{\prime}\rangle/\sqrt{\langle u^{\prime 2}\rangle\langle b^{\prime 2}\rangle}$ and $\langle\mathbf{u}^{\prime}\cdot\mathbf{w}^{\prime}\rangle/\sqrt{\langle u^{\prime 2}\rangle\langle w^{\prime 2}\rangle}$ are plotted against time and they are both clearly non-zero in all the considered cases; the cross-helicity is plotted in red and the blue lines correspond to the kinetic helicity whereas their time averages are marked with white lines. In addition, only for the sake of reference, the figures also show the estimates of the non-equilibrium effect in the form
\begin{equation}
\alpha_{\rm neq}\approx -\frac{1}{3}\frac{\langle\mathbf{u}^{\prime}\cdot\mathbf{b}^{\prime}\rangle}{\sqrt{\langle u^{\prime 2}\rangle\langle b^{\prime 2}\rangle}}
\int_{-\infty}^{\tau}\mathrm{d}\tau_{1}\left[\langle\mathbf{u}^{\prime}\left(\mathbf{x},\tau\right)\cdot\mathbf{j}^{\prime}\left(\mathbf{x},\tau_1\right)\rangle - \langle\mathbf{u}^{\prime}\left(\mathbf{x},\tau_1\right)\cdot\mathbf{j}^{\prime}\left(\mathbf{x},\tau\right)\rangle\right],
\label{est_alpha_neq}
\end{equation}
which can be compared with the following standard estimate of the $\alpha$-effect, associated with the presence of the kinetic and current helicities
\begin{equation}
\alpha_{S}\approx -\frac{1}{3}\tau_t
\left(\langle\mathbf{u}^{\prime}\cdot\mathbf{w}^{\prime}\rangle - \langle\mathbf{b}^{\prime}\cdot\mathbf{j}^{\prime}\rangle\right),
\label{est_alpha_standard}
\end{equation}
where $\tau_t=1/u_{\rm rms}k_{\rm f}$ is the turnover time of most energetic turbulent eddies, with $u_{\rm rms}=\sqrt{\langle u'^2\rangle}$ and $k_{\rm f}=30\,k_1$ denoting the
forcing the wavenumber ($k_1=2\pi/L$ is the wavenumber of the box of length $L$).
%
%
%
%
%

Although in the numerically studied cases the statistical non-stationarity of turbulence is rather weak and the estimate of the $\alpha_{\rm neq}$ coefficient is always at least an order of magnitude weaker than $\alpha_{\rm S}$, the former is clearly different from zero and its relative importance seems to correlate with the magnitude of the cross-helicity.
The relative enhancement of the $\alpha_{\rm neq}$-effect visible
for a stronger magnetic field (Run~B) and weaker gravity (Run~E)
corresponds to the enhancement of the cross-helicity
with respect to the kinetic one.
Of course in the latter case (see figure~\ref{fig:b0-1}),
although the $\alpha_{\rm neq}$ coefficient has the largest relative
magnitude it also has a different sign than $\alpha_{\rm S}$, hence
in this case the non-equilibrium effects tend to suppress the standard
dynamo effect.
In figure~\ref{fig:b0-01_diff_g} we see, that weak magnetic field and
strong gravity have suppressed the non-equilibrium effect to a very
small relative magnitude.

\begin{table}
  \begin{center}
\def~{\hphantom{0}}
  \begin{tabular}{lcccccccccc}
& $\frac{g}{c_{\rm s}^2 k_1}$ & $\frac{v_{A0}}{c_{\rm s}}$ & 
$\frac{\langle\mathbf{u}^{\prime}\cdot\mathbf{b}^{\prime}\rangle}{\sqrt{\langle u^{\prime 2}\rangle\langle b^{\prime 2}\rangle}}$ &
$\frac{\langle\mathbf{u}^{\prime}\cdot\mathbf{w}^{\prime}\rangle}{\sqrt{\langle u^{\prime 2}\rangle\langle w^{\prime 2}\rangle}}$ &
$\frac{\langle\mathbf{b}^{\prime}\cdot\mathbf{j}^{\prime}\rangle}{\sqrt{\langle u^{\prime 2}\rangle\langle w^{\prime 2}\rangle}}$ &
$\frac{\alpha_{\rm neq}}{\alpha_0}$ &
$\frac{\alpha_{\rm S}}{\alpha_0}$ &
$\frac{u_{\rm rms}}{c_{\rm s}}$ &
$\frac{v_{A}^{\rm rms}}{c_{\rm s}}$ &
\\[7pt]
C & 0.5 &  0.01 & $-9.8\times10^{-3}$ & $-1.6\times10^{-2}$ & $-2.0\times10^{-4}$ & $7.8\times10^{-4}$ & $1.8\times10^{-2}$ & 0.10 & 0.03 \\
A & 1.0 &  0.01 & $-1.7\times10^{-2}$ & $-3.0\times10^{-2}$ & $-3.3\times10^{-4}$ & $1.1\times10^{-3}$ & $3.5\times10^{-2}$ & 0.11 & 0.04 \\
D & 2.0 &  0.01 & $-2.0\times10^{-2}$ & $-3.6\times10^{-2}$ & $-2.8\times10^{-4}$ & $6.1\times10^{-4}$ & $4.1\times10^{-2}$ & 0.16 & 0.04 \\
E & 0.5 &  0.10 & $-5.5\times10^{-2}$ & $-1.9\times10^{-2}$ & $-6.2\times10^{-4}$ & $-5.6\times10^{-3}$ & $1.5\times10^{-2}$ & 0.08 & 0.07 \\
B & 1.0 &  0.10 & $-5.3\times10^{-2}$ & $-3.2\times10^{-2}$ & $-1.2\times10^{-2}$ & $2.3\times10^{-3}$ & $1.8\times10^{-2}$ & 0.09 & 0.12 \\
  \end{tabular}
  \caption{Summary of the simulation results for Runs~A--E.}
  \label{tab:summary}
  \end{center}
\end{table}

\begin{table}
  \begin{center}
\def~{\hphantom{0}}
  \begin{tabular}{lcccccccccc}
& $k_{\rm f}$ &
$\frac{\langle\mathbf{u}^{\prime}\cdot\mathbf{b}^{\prime}\rangle}{\sqrt{\langle u^{\prime 2}\rangle\langle b^{\prime 2}\rangle}}$ &
$\frac{\langle\mathbf{u}^{\prime}\cdot\mathbf{w}^{\prime}\rangle}{\sqrt{\langle u^{\prime 2}\rangle\langle w^{\prime 2}\rangle}}$ &
$\frac{\langle\mathbf{b}^{\prime}\cdot\mathbf{j}^{\prime}\rangle}{\sqrt{\langle u^{\prime 2}\rangle\langle w^{\prime 2}\rangle}}$ &
$\frac{\alpha_{\rm neq}}{\alpha_0}$ &
$\frac{\alpha_{\rm S}}{\alpha_0}$ &
$\frac{u_{\rm rms}}{c_{\rm s}}$ &
$\frac{v_{A}^{\rm rms}}{c_{\rm s}}$ &
\\[7pt]
A & 30 & $-1.7\times10^{-2}$ & $-3.0\times10^{-2}$ & $-3.3\times10^{-4}$ & $1.1\times10^{-3}$ & $3.5\times10^{-2}$ & 0.11 & 0.04 \\
A2& 10 & $-1.3\times10^{-1}$ & $-1.2\times10^{-1}$ & $1.3\times10^{-3}$ & $-1.7\times10^{-2}$ & $6.9\times10^{-2}$ & 0.12 & 0.12 \\
A3&  3 & $-6.4\times10^{-2}$ & $-2.1\times10^{-1}$ & $-3.0\times10^{-2}$ & $-6.0\times10^{-3}$ & $5.5\times10^{-2}$ & 0.19 & 0.09 \\
  \end{tabular}
  \caption{Summary of the simulation results for Runs~A, A2, and A3.}
  \label{tab:summary2}
  \end{center}
\end{table}

At smaller scale separation, i.e., for smaller values of $k_{\rm f}$,
we expect the turbulence to be more intermittent and degree of
non-stationarity to be enhanced.
To address this possibility, we have performed additional simulations
for smaller values of $k_{\rm f}$ with the other parameters being the
same as for Run~A.
The results shown in table~2 do show that $\alpha_{\rm neq}$ is twice
as large when $k_{\rm f}$ is reduced from 30 to 10, but an additional
decrease of $k_{\rm f}$ from 10 to 3 does not lead to an additional
increase of $\alpha_{\rm neq}$.
To some extent, however, this is caused by the normalization by
$\alpha_0$, which has increased by about 60\%.

We conclude, that in fully developed helical turbulence, that is in turbulence with strong kinetic helicity, the cross-helicity is rather likely to be produced as well and at least for some periods of time the two helicities can coexist.

\section{Conclusions}

We have analysed the hydromagnetic dynamo process in non-equilibrium turbulence. It was shown that in non-equilibrium MHD turbulence the effect of the infinitesimal-impulse cross responses $\mathbf{u}'\leftrightarrow\mathbf{b}'$ is pronounced, which vanishes in stationary state. This creates additional terms in the expression for the large-scale electromotive force.

The main conclusion is that the non-equilibrium effects in MHD turbulence modify the $\alpha$-effect by introducing a correction dependent on the square of the non-dimensional cross-helicity $\Upsilon = \langle\mathbf{u}'\cdot\mathbf{b}'\rangle/\sqrt{\langle u^{\prime2}\rangle \langle b^{\prime2}\rangle}$, the kinetic helicity and their history in the MHD turbulence, which takes  the form provided in (\ref{eq:alpha_neq_est_fin}).
This requires coexistence of both, the kinetic and cross-helicities in the turbulent flow. The discussion of the production mechanisms of the cross-helicity, provided in section \ref{subsec:coexist} and the results of numerical simulations, lead to a conclusion that such coexistence is possible and perhaps even ubiquitous in many natural systems. Simple strong production mechanisms have been identified already and thoroughly discussed in earlier works, e.g.\ Yokoi and Hoshino (2011).

The non-equilibrium effects in turbulence affect also other components of the mean EMF (\ref{eq:EMF_gen}), that is the turbulent diffusivity $\beta$ and the coefficients $\zeta$ and $\gamma$ in a non-trivial way, through the effect of the Green's cross-response functions $G_{ub}$ and $G_{bu}$. This interesting topic should be investigated in more detail in future studies.

~

\section*{Acknowledgements}

We would like to thank the Isaac Newton Institute for Mathematical
Sciences, Cambridge, for support and hospitality during the programme
``Frontiers in dynamo theory: from the Earth to the stars'' (DYT2) where much
of the work on this paper was undertaken.

\section*{Funding}

KAM was supported by a subsidy from the Polish
Ministry of Education and Science for the Institute of
Geophysics, Polish Academy of Sciences. NY was supported by the Japan Society of the Promotion of Science (JSPS) Grants-in-Aid for Scientific Research JP18H01212.
We also acknowledge the support of the EPSRC grant no EP/R014604/1 and
the Swedish Research Council (Vetenskapsr{\aa}det, 2019-04234).
Nordita is sponsored by Nordforsk.
We acknowledge the allocation of computing resources provided by the
Swedish National Allocations Committee at the Center for Parallel
Computers at the Royal Institute of Technology in Stockholm and
Link\"oping.

\section*{Declaration of Interests}
The authors report no conflict of interest.

\section*{Data availability statement}
The data that support the findings of this study are openly available
on Zenodo at doi:10.5281/zenodo.7683615 (v2023.02.28).
All calculations have been performed with the {\sc Pencil Code};
DOI:10.5281/zenodo.3961647.

\section*{Author ORCID}

\noindent
K. A. Mizerski, https://orcid.org/0000-0003-0106-675X\\
N. Yokoi, https://orcid.org/0000-0002-5242-7634\\
A. Brandenburg, https://orcid.org/0000-0002-7304-021X

\appendix

\section{Outline of the two-scale direct-interaction approximation (TSDIA) with self- and cross-interaction response functions for the velocity and magnetic fields.}\label{appA}

The two-scale direct-interaction approximation (TSDIA) is a combination of the direct-interaction approximation (DIA) for strongly nonlinear homogeneous isotropic turbulence and the multiple-scale analysis with the derivative expansion with respect to the large-scale inhomogeneity. The TSDIA provides a powerful tool for investigating strongly-nonlinear turbulence with large-scale inhomogeneities. In applying the TSDIA scheme to the magnetohydrodynamic turbulence, the Els\"{a}sser variable formulation has been often adopted. In this formulation, symmetries of the velocity and magnetic-field equations are fully utilized, which reduces the complexities in treating the original MHD equations. The correspondence between the Els\"{a}sser variable formulation and the usual velocity--magnetic-field formulation in the TSDIA has been discussed in some literature (Yoshizawa 1998, Hamba \& Sato 2008, Yokoi 2013). Here, we present the outline of the TSDIA formulation under the velocity and magnetic-field variables with special references to the self- and cross-interaction response functions in the MHD turbulence. For the outline of the DIA in the context of the TSDIA, the reader is referred to textbooks such as Yoshizawa (1998) and Yokoi (2020).

\paragraph{Wave-number space equations}
We introduce the Fourier representation concerning the fast space variable $\mbox{\boldmath$\xi$}$ as
\begin{equation}
	f'(\mbox{\boldmath$\xi$},{\bf{X}};\tau,T)
	= \int d{\bf{k}} f({\bf{k}},{\bf{X}};\tau,T) 
		\exp[- i {\bf{k}} \cdot 
		(\mbox{\boldmath$\xi$} - \langle{\bf{U}}\rangle \tau)],
		\label{eq:A_fourier_rep}
\end{equation}
where the Fourier transform of the fast variable is taken in the frame co-moving with the local mean velocity $\langle{\bf{U}}\rangle$. Hereafter, for the sake of simplicity of notation, the arguments of the slow variable for the fluctuation field $f(\mbox{\boldmath$\xi$},{\bf{X}};\tau,T)$ is suppressed and just denoted as $f(\mbox{\boldmath$\xi$};\tau)$.

The system of two-scale differential equations under the velocity and magnetic-field variables in the wavenumber space is written as
\begin{eqnarray}
	&&\frac{\partial u^i({\bf{k}};\tau)}{\partial \tau}
	+ \nu k^2 u^i({\bf{k}};\tau)+ i k^j \langle B\rangle^j b^i({\bf{k}};\tau)
	\nonumber\\
	&&- i M^{ij\ell}({\bf{k}}) \iint d{\bf{p}} d{\bf{q}}\ 
		\delta({\bf{k}} - {\bf{p}} - {\bf{q}}) \times \left[ {
		u^j({\bf{p}};\tau) u^\ell({\bf{q}};\tau)
		- b^j({\bf{p}};\tau) b^\ell({\bf{q}};\tau)
	} \right] 
	\nonumber\\
	&&\hspace{80pt} =\delta \left[ {
	- D^{ij}({\bf{k}}) 
	\frac{\widehat{D} u^j({\bf{k}};\tau)}{D T_{\rm{I}}}
	- D^{ij}({\bf{k}}) u^m({\bf{k}};\tau) \left( {
		\frac{\partial \langle U\rangle^j}{\partial X^m}
		+ \epsilon^{mj\ell} \Omega_0^\ell
		} \right)
	} \right.
	\nonumber\\
	&& \hspace{80pt} \left. { \hspace{22pt}
	+ \langle B\rangle^j \frac{\partial b^i({\bf{k}};\tau)}{\partial X_{\rm{I}}^j}
	+D^{ij}({\bf{k}}) b^m({\bf{k}};\tau) \frac{\partial \langle B\rangle^j}{\partial X^m}
	} \right],
		\label{eq:u(k)_eq}
\end{eqnarray}
\begin{equation}
	- i k^ j u^j({\bf{k}};\tau)
	+ \delta \frac{\partial u^j({\bf{k}};\tau)}{\partial X^j}
	= 0,
		\label{eq:A_u_sol_eq}
\end{equation}
\begin{eqnarray}
	&&\frac{\partial b^i({\bf{k}};\tau)}{\partial \tau}
	+ \eta k^2 b^i({\bf{k}};\tau)+ i k^j \langle B\rangle^j u^i({\bf{k}};\tau)
	\nonumber\\
	&&+ i N^{ij\ell}({\bf{k}}) \iint d{\bf{p}} d{\bf{q}}\ 
		\delta({\bf{k}} - {\bf{p}} - {\bf{q}}) \times
	\left[ {
		b^j({\bf{p}};\tau) u^\ell({\bf{q}};\tau)
		- u^j({\bf{p}};\tau) b^\ell({\bf{q}};\tau)
	} \right]
	\nonumber\\
	&&\hspace{80pt} = \delta \left[ {
		- D^{ij}({\bf{k}}) 
		\frac{\widehat{D} b^j({\bf{k}};\tau)}{D T_{\rm{I}}}
		+ D^{ij}({\bf{k}}) b^m({\bf{k}};\tau) \left( {
		\frac{\partial \langle U\rangle^j}{\partial X^m}
	+ \epsilon^{mj\ell} \Omega_0^\ell
	} \right)
	} \right.
	\nonumber\\
	&&  \hspace{80pt}\left. { \hspace{22pt}
	+ \langle B\rangle^j \frac{\partial u^i({\bf{k}};\tau)}{\partial X_{\rm{I}}^j}
	- D^{ij}({\bf{k}}) u^m({\bf{k}};\tau) \frac{\partial \langle B\rangle^j}{\partial X^m}
	} \right],
	\label{eq:A_b(k)_eq}
\end{eqnarray}
\begin{equation}
	- i k^ j b^j({\bf{k}};\tau)
	+ \delta \frac{\partial b^j({\bf{k}};\tau)}{\partial X^j}
	= 0,
		\label{eq:A_b_sol_eq}
\end{equation}
where
\begin{equation}
	\left( {\nabla_{{\bf{X}}{\rm{I}}}, \frac{D}{DT_{\rm{I}}}}  \right)
	= \exp \left( {-i {\bf{k}} \cdot {\bf{U}}\tau} \right) 
		\left( {\nabla_{\bf{X}}, \frac{D}{DT}} \right)
		\exp \left( {i {\bf{k}} \cdot {\bf{U}} \tau} \right)
		\label{eq:A_int_rep}
\end{equation}
is the differential operators in the interaction representation. Here in (\ref{eq:u(k)_eq}) and (\ref{eq:A_b(k)_eq}),
\begin{equation}
	M^{ijk}({\bf{k}})
	= \frac{1}{2}\left(k^j D^{ik}({\bf{k}}) + k^k D^{ij}({\bf{k}})\right),
		\label{eq:A_Mijk_def}
\end{equation}
with the solenoidal projection operator
\begin{equation}
	D^{ij}({\bf{k}}) = \delta^{ij} - \frac{k^i k^j}{k^2},
		\label{eq:A_sol_proj_op}
\end{equation}
and
\begin{equation}
	N^{ijk}({\bf{k}}) = k^j \delta^{ik} - k^k \delta^{ij}.
		\label{eq:A_Nijk_def}
\end{equation}
The operators $M$ and $N$ are point vertices showing the wave-number conservation among the nonlinear mode coupling with $\delta({\bf{k}}-{\bf{p}}-{\bf{q}})$.

In (\ref{eq:u(k)_eq}) and (\ref{eq:A_b(k)_eq}), to keep the material derivatives objective (invariant with respect to rotations), we adopt a co-rotational derivative
\begin{equation}
	\frac{\widehat{D} u'{}^i}{DT}
	= \frac{\partial u'{}^i}{\partial T}
	+ \langle U\rangle^j \frac{\partial u'{}^i}{\partial X^j}
	+ \epsilon^{jik} \Omega_0^k u'{}^j
		\label{eq:A_corot_der}
\end{equation}
with
\begin{equation}
	\mbox{\boldmath$\Omega$}_0
	= \mbox{\boldmath$\Omega$}/\delta
		\label{eq:A_Omega0_def}
\end{equation}

in place of the Lagrange or advective derivative
\begin{equation}
	\frac{Du'{}^i}{DT}
	= \frac{\partial u'{}^i}{\partial t}
	+ \langle U\rangle^j \frac{\partial u'{}^i}{\partial x^j},
\end{equation}
which is not objective with respect to a rotation.

\paragraph{Scale-parameter expansion}
	We expand a field $f({\bf{k}};\tau)$ with respect to the scale parameter $\delta$, and further expand each field by the external field (the mean magnetic field in the present case) as
\begin{eqnarray}
	f^i({\bf{k}};\tau)
	&=& \sum_{n=0}^{\infty} 
	\delta^n  f_{n}^i({\bf{k}};\tau)
	- \sum_{n=0}^{\infty}
	\delta^{n+1} i \frac{k^i}{k^2} 
	\frac{\partial}{\partial X_{\rm{I}}^{j}} f_{n}^j({\bf{k}};\tau)
	\nonumber\\
	&=& \sum_{n=0}^{\infty} \sum_{m=0}^{\infty}
	\delta^n  f_{nm}^i({\bf{k}};\tau)
	- \sum_{n=0}^{\infty} \sum_{m=0}^{\infty}
	\delta^{n+1} i \frac{k^i}{k^2} 
	\frac{\partial}{\partial X_{\rm{I}}^{j}} f_{nm}^j({\bf{k}};\tau).
	\label{eq:A_scale_exps}
\end{eqnarray}
In this two-scale formulation, inhomogeneities and anisotropies enter with the scale parameter $\delta$ and the external parameters $\langle{\bf{B}}\rangle$ in higher-order fields. The lowest-order fields $f_{00}$ fields correspond to the homogeneous and isotropic turbulence.

Using the expansion (\ref{eq:A_scale_exps}), we write the equations of each order in matrix form. With the abbreviated form of the spectral integral
\begin{equation}
	\int_\Delta
	= \iint d{\bf{p}} d{\bf{q}}\ \delta({\bf{k}} - {\bf{p}} - {\bf{q}}),
	\label{eq:A_abbre_pq_int}
\end{equation}
the $f_{00}({\bf{k}};\tau)$ equations are given as
\begin{eqnarray}
	&\left( {
	\begin{array}{c}
		0\\
		0\rule{0.ex}{5.ex}
	\end{array}
	} \right)={\everymath{\displaystyle}\left( {
	\begin{array}{cc}
		\frac{\partial}{\partial \tau} + \nu k^2 & 0\\
		0 & \frac{\partial}{\partial \tau} + \eta k^2
	\end{array}
	} \right)}
	\left( {
	\begin{array}{cc}
		u_{00}^{i}({\bf{k}};\tau)\\
		b_{00}^{i}({\bf{k}};\tau)\rule{0.ex}{5.ex}
	\end{array}
	} \right)\hspace{5cm}
	\nonumber\\
	& \hspace{-5mm}{\everymath{\displaystyle}
	+ i \left( {
	\begin{array}{cc}
		- M^{ij\ell}({\bf{k}}) \int_\Delta u_{00}^j({\bf{p}};\tau) 
		& M^{ij\ell}({\bf{k}}) \int_\Delta b_{00}^j({\bf{p}};\tau)\\
		N^{ij\ell}({\bf{k}}) \int_\Delta b_{00}^j({\bf{p}};\tau) 
		& - N^{ij\ell}({\bf{k}}) \int_\Delta u_{00}^j({\bf{p}};\tau)
	\end{array}
	} \right)
	\left( {
	\begin{array}{c}
		u_{00}^{\ell}({\bf{q}};\tau)\\
		b_{00}^{\ell}({\bf{q}};\tau)\rule{0.ex}{5.ex}
	\end{array}
	} \right)},
	\label{eq:A_f00_eqs}
\end{eqnarray}
the $f_{01}({\bf{k}};\tau)$ equations are given as
\begin{eqnarray}
	&&{\everymath{\displaystyle}\left( {
	\begin{array}{cc}
		\frac{\partial}{\partial \tau} + \nu k^2 & 0\\
		0 & \frac{\partial}{\partial \tau} + \eta k^2
	\end{array}
	} \right)}
	\left( {
	\begin{array}{cc}
		u_{01}^{i}({\bf{k}};\tau)\\
		b_{01}^{i}({\bf{k}};\tau)\rule{0.ex}{5.ex}
	\end{array}
	} \right)
	\nonumber\\
	&& {\everymath{\displaystyle}
	+ i \left( {
	\begin{array}{cc}
		- 2M^{ij\ell}({\bf{k}}) \int_\Delta u_{00}^j({\bf{p}};\tau) 
		& 2M^{ij\ell}({\bf{k}}) \int_\Delta b_{00}^j({\bf{p}};\tau)\\
		N^{ij\ell}({\bf{k}}) \int_\Delta b_{00}^j({\bf{p}};\tau) 
		& - N^{ij\ell}({\bf{k}}) \int_\Delta u_{00}^j({\bf{p}};\tau)
	\end{array}
	} \right)
	\left( {
	\begin{array}{c}
		u_{01}^{\ell}({\bf{q}};\tau)\\
		b_{01}^{\ell}({\bf{q}};\tau)\rule{0.ex}{5.ex}
	\end{array}
	} \right)}
	\nonumber\\
	&& \hspace{3cm}= - i k^j \langle B\rangle^j 
	\left( {
	\begin{array}{cc}
		0 & 1\\
		1 & 0 \rule{0.ex}{5.ex}
	\end{array}
	} \right)
	\left( {
	\begin{array}{c}
		u_{00}^i({\bf{k}};\tau)\\
		b_{00}^i({\bf{k}};\tau)\rule{0.ex}{5.ex}
	\end{array}
	} \right)
	\equiv \left( {
	\begin{array}{c}
		F_{01u}^i\\
		F_{01b}^i\rule{0.ex}{5.ex}
	\end{array}
	} \right),\qquad\qquad\qquad
		\label{eq:A_f01_eqs}
\end{eqnarray}
and the $f_{10}({\bf{k}};\tau)$ equations are
\begin{align}
	&{\everymath{\displaystyle}\left( {
	\begin{array}{cc}
		\frac{\partial}{\partial \tau} + \nu k^2 & 0\\
		0 & \frac{\partial}{\partial \tau} + \eta k^2
	\end{array}
	} \right)}
	\left( {
	\begin{array}{cc}
		u_{10}^{i}({\bf{k}};\tau)\\
		b_{10}^{i}({\bf{k}};\tau)\rule{0.ex}{5.ex}
	\end{array}
	} \right)
	\nonumber\\
	& {\everymath{\displaystyle}
	+ i \left( {
	\begin{array}{cc}
		- 2M^{ij\ell}({\bf{k}}) \int_\Delta u_{00}^j({\bf{p}};\tau) 
		& 2M^{ij\ell}({\bf{k}}) \int_\Delta b_{00}^j({\bf{p}};\tau)\\
		N^{ij\ell}({\bf{k}}) \int_\Delta b_{00}^j({\bf{p}};\tau) 
		& - N^{ij\ell}({\bf{k}}) \int_\Delta u_{00}^j({\bf{p}};\tau)
	\end{array}
	} \right)
	\left( {
	\begin{array}{c}
		u_{10}^{\ell}({\bf{q}};\tau)\\
		b_{10}^{\ell}({\bf{q}};\tau)\rule{0.ex}{5.ex}
	\end{array}
	} \right)}
	\nonumber\\
	& \hspace{0.5cm}= {\everymath{\displaystyle} \langle B\rangle^j \frac{\partial}{\partial X_{\rm{I}}^j}
	\left( {
	\begin{array}{cc}
		0
		& 1\\
		1
		& 0 \rule{0.ex}{5.ex}
	\end{array}
	} \right)
	\left( {
	\begin{array}{c}
		u_{00}^i({\bf{k}})\\
		b_{00}^i({\bf{k}})\rule{0.ex}{5.ex}
	\end{array}
	} \right)}
	- D^{ij}({\bf{k}}) \frac{\widehat{D}}{DT_{\rm{I}}} {\everymath{\displaystyle}\left( {
	\begin{array}{cc}
		1
		& 0\\
		0
		& 1 \rule{0.ex}{5.ex}
	\end{array}
	} \right)
	\left( {
	\begin{array}{c}
		u_{00}^j({\bf{k}})\\
		b_{00}^j({\bf{k}})\rule{0.ex}{5.ex}
	\end{array}
	} \right)}
	\nonumber\\
	& \hspace{1cm}+ {\everymath{\displaystyle}\left( {
	\begin{array}{cc}
		- D^{ij}({\bf{k}}) \left( {
		\frac{\partial \langle U\rangle^j}{\partial X^\ell}
		+ \epsilon^{\ell jn} \Omega_0^n
		} \right)
		& D^{ij}({\bf{k}}) \frac{\partial \langle B\rangle^j}{\partial X^\ell}\\
		- D^{ij}({\bf{k}}) \frac{\partial \langle B\rangle^j}{\partial X^\ell}
		& D^{ij}({\bf{k}}) \left( {
		\frac{\partial \langle U\rangle^j}{\partial X^\ell} 
		+ \epsilon^{\ell jn} \Omega_0^n
		} \right)
	\end{array}
	} \right)
	\left( {
	\begin{array}{c}
		u_{00}^\ell({\bf{k}};\tau)\\
		b_{00}^\ell({\bf{k}};\tau)\rule{0.ex}{5.ex}
	\end{array}
	} \right)}
	\nonumber\\
	& \hspace{0.5cm}\equiv \left( {
	\begin{array}{c}
		F_{10u}^i\\
		F_{10b}^i \rule{0.ex}{5.ex}
	\end{array}
	} \right),
		\label{eq:A_f10_eqs}
	\end{align}
where, $F_{01u}$, $F_{01b}$, $F_{10u}$, and $F_{10b}$ denote each component of the second right-hand sides (r.h.s.) of (\ref{eq:A_f01_eqs}) and (\ref{eq:A_f10_eqs}). They can be regarded as the forcing for the evolution equations of $f_{01}({\bf{k}};\tau)$ and $f_{10}({\bf{k}};\tau)$, respectively.

\paragraph{Introduction of Green's functions}
For the purpose of solving these differential equations, we introduce the Green's functions associated with (\ref{eq:A_f00_eqs}). We consider the response of the turbulence to an infinitesimal disturbance. Reflecting the structure of the MHD equations and the field expansion (\ref{eq:A_scale_exps}), the left-hand side of the  linearized differential equations for the Green's function is in the same form as the l.h.s.\ of (\ref{eq:A_f01_eqs}) and (\ref{eq:A_f10_eqs}) or the differential operators to the $f_{01}({\bf{k}};\tau)$ and $f_{10}({\bf{k}};\tau)$ fields. In order to treat mutual interaction between the velocity and magnetic field, we consider four Green's functions; the Green function $G_{uu}$ representing the response of the velocity field ${\bf{u}}$ to the velocity perturbation ${\bf{u}}$, $G_{ub}$ the response of ${\bf{u}}$ to the magnetic perturbation ${\bf{b}}$, $G_{bu}$ the response of ${\bf{b}}$ to the velocity perturbation ${\bf{u}}$, and $G_{bb}$ the response of magnetic field ${\bf{b}}$ to the magnetic perturbation ${\bf{b}}$. From the l.h.s.\ of (\ref{eq:A_f01_eqs}) and (\ref{eq:A_f10_eqs}) we construct the system of equations representing the responses to the infinitesimal forcing. It follows that these four Green's functions should be defined by their evolution equations as
\begin{align}
	&{\everymath{\displaystyle}\left( {
	\begin{array}{cc}
		\frac{\partial}{\partial \tau} + \nu k^2 & 0\\
		0 & \frac{\partial}{\partial \tau} + \eta k^2
	\end{array}
	} \right)}
	\left( {
	\begin{array}{cc}
		G_{uu}^{ij} & G_{bu}^{ij}\\
		G_{ub}^{ij} & G_{bb}^{ij}\rule{0.ex}{5.ex}
	\end{array}
	} \right)
	\nonumber\\
	& {\everymath{\displaystyle}
	+ i \left( {
	\begin{array}{cc}
		- 2M^{ikm} \int_\Delta u_{00}^k & 2M^{ikm} \int_\Delta b_{00}^k\\
		N^{ikm} \int_\Delta b_{00}^k & - N^{ikm} \int_\Delta u_{00}^k
	\end{array}
	} \right)
	\left( {
	\begin{array}{cc}
		G_{uu}^{mj} & G_{bu}^{mj}\\
		G_{ub}^{mj} & G_{bb}^{mj}\rule{0.ex}{5.ex}
	\end{array}
	} \right)}
	= \delta^{ij} \delta(\tau - \tau') \left( {
	\begin{array}{cc}
		1 & 0\\
		0 & 1\rule{0.ex}{5.ex}
	\end{array}
	} \right).
		\label{eq:A_G_eqs}
\end{align}
Considering that the r.h.s.\ of (\ref{eq:A_f01_eqs}) and (\ref{eq:A_f10_eqs}) are the force terms, we formally solve $f_{01}$ and $f_{10}$ fields with the aid of the Green's functions. The $f_{01}$ fields are expressed as
\begin{equation}
	\left( {
	\begin{array}{c}
		u_{01}^i\\
		b_{01}^i \rule{0.ex}{5.ex}
	\end{array}
	} \right)
	= \int_{-\infty}^\tau \!\!\! d\tau_1 
	\left( {
	\begin{array}{cc}
		G_{uu}^{ij} & G_{ub}^{ij}\\
		G_{bu}^{ij} & G_{bb}^{ij} \rule{0.ex}{5.ex}
	\end{array}
	} \right)
	\left( {
	\begin{array}{c}
		F_{01u}^j\\
		F_{01b}^j \rule{0.ex}{5.ex}
	\end{array}
	} \right).
		\label{eq:A_f01_sol}
\end{equation}
Note that ${\bf{u}}_{01}$ and ${\bf{b}}_{01}$ are expressed the ${\bf{b}}_{00}$ and ${\bf{u}}_{00}$ coupled with the mean magnetic field $\langle{\bf{B}}\rangle$, respectively. As this result, ${\bf{u}}_{01}$ and ${\bf{b}}_{01}$ multiplied by ${\bf{b}}_{00}$ and ${\bf{u}}_{00}$ in an external product manner will not contribute to the EMF.

On the other hand, the $f_{10}$ fields are expressed as
\begin{equation}
	\left( {
	\begin{array}{c}
		u_{10}^i\\
		b_{10}^i \rule{0.ex}{5.ex}
	\end{array}
	} \right)
	= \int_{-\infty}^\tau \!\!\! d\tau_1 
	\left( {
	\begin{array}{cc}
		G_{uu}^{ij} & G_{ub}^{ij}\\
		G_{bu}^{ij} & G_{bb}^{ij} \rule{0.ex}{5.ex}
	\end{array}
	} \right)
	\left( {
	\begin{array}{c}
		F_{10u}^j\\
		F_{10b}^j \rule{0.ex}{5.ex}
	\end{array}
	} \right).
		\label{eq:A_f10_sol}
\end{equation}

\paragraph{Statistical assumption on the basic fields}
	We assume that the basic or lowest-order fields are homogeneous and isotropic. 
\begin{equation}
	\frac{\left\langle {
		\vartheta_{00}^i({\bf{k}};\tau) \chi_{00}^j({\bf{k}}';\tau')
		} \right\rangle}{\delta({\bf{k}} + {\bf{k}}')}
	= D^{ij}({\bf{k}}) Q_{\vartheta\chi}({\bf{k}};\tau,\tau')
	+ \frac{i}{2} \frac{k^\ell}{k^2} \epsilon^{ij\ell}
		H_{\vartheta\chi}({\bf{k}};\tau,\tau'),
		\label{eq:A_basic_fld_assum}
\end{equation}
where $\mbox{\boldmath$\vartheta$}_{00}$ and $\mbox{\boldmath$\chi$}_{00}$ represent one of ${\bf{u}}_{00}$ and ${\bf{b}}_{00}$, and the indices $\vartheta$ and $\chi$ do one of $u$ and $b$. The Green's functions are written as
\begin{equation}
	\langle {G_{\vartheta\chi}^{ij}({\bf{k}};\tau,\tau')} \rangle
	= D^{ij}({\bf{k}}) G_{\vartheta\chi}({\bf{k}};\tau,\tau').
		\label{eq:A_G_assum}
\end{equation}

The spectral functions, $Q_{uu}$, $Q_{bb}$, $Q_{ub}$, $H_{uu}$, $H_{bb}$, $H_{ub}$, and $H_{bu}$, are related to the turbulent statistical quantities (the turbulent kinetic energy, magnetic energy, cross helicity, kinetic helicity, electric-current helicity, torsional correlations between velocity and magnetic field) of the basic or lowest-order fields as
\begin{equation}
	\int d{\bf{k}}\ Q_{uu}(k;\tau,\tau) 
	= \langle {{\bf{u}}'_{00}{}^2} \rangle/2,
		\label{eq:A_Quu}
\end{equation}
\begin{equation}
	\int d{\bf{k}}\ Q_{bb}(k;\tau,\tau) 
	= \langle {{\bf{b}}'_{00}{}^2} \rangle/2,
		\label{eq:A_Qbb}
\end{equation}
\begin{equation}
	\int d{\bf{k}}\ Q_{ub}(k;\tau,\tau) 
	= \langle {{\bf{u}}'_{00} \cdot {\bf{b}}'_{00}} \rangle,
		\label{eq:A_Qub}
\end{equation}
\begin{equation}
	\int d{\bf{k}}\ H_{uu}(k;\tau,\tau) 
	= \langle {{\bf{u}}'_{00} \cdot \mbox{\boldmath$\omega$}'_{00}} \rangle,
		\label{eq:A_Huu}
\end{equation}
\begin{equation}
	\int d{\bf{k}}\ H_{bb}(k;\tau,\tau) 
	= \langle {{\bf{b}}'_{00} \cdot {\bf{j}}'_{00}} \rangle,
		\label{eq:A_Hbb}
\end{equation}
\begin{equation}
	\int d{\bf{k}}\ H_{ub}(k;\tau,\tau) 
	= \langle {{\bf{u}}'_{00} \cdot {\bf{j}}'_{00}} \rangle,
		\label{eq:A_Hub}
\end{equation}
\begin{equation}
	\int d{\bf{k}}\ H_{bu}(k;\tau,\tau) 
	= \langle {{\bf{b}}'_{00} \cdot \mbox{\boldmath$\omega$}'_{00}} \rangle.
	\label{eq:A_Hbu}
\end{equation}

\paragraph{Calculation of the electromotive force (EMF)}
The turbulent electromotive force (EMF) is expressed in terms of the wave-number representation of the velocity and magnetic-field as
\begin{equation}
	E_{\rm{M}}^i
	\equiv \epsilon^{ijk} \langle {u'{}^j b'{}^k} \rangle
	= \epsilon^{ijk} \int d{\bf{k}}\ 
		\langle {u^j({\bf{k}};\tau) b^k({\bf{k}}';\tau)} \rangle 
		/ \delta({\bf{k}} + {\bf{k}}').
		\label{eq:A_emf_def}
\end{equation}
Using the results of (\ref{eq:A_f01_sol}) and (\ref{eq:A_f10_sol}), we calculate the velocity--magnetic-field correlation up to the $f_{01}g_{00}$ and $f_{10} g_{00}$ orders as
\begin{equation}
	\langle {u^j b^k} \rangle
	= \langle {u_{00}^j b_{00}^k} \rangle
	+ \langle {u_{01}^j b_{00}^k} \rangle 
	+ \langle {u_{00}^j b_{01}^k} \rangle
	+ \delta \langle {u_{10}^j b_{00}^k} \rangle
	+ \delta \langle {u_{00}^j b_{10}^k} \rangle
	+ \cdots.
		\label{eq:A_ub_expansion}
\end{equation}

In the direct-interaction approximation (DIA) formalism, the lowest-order spectral functions $Q_{uu}$, $Q_{bb}$, $Q_{ub}$, $H_{uu}$, $H_{bb}$, $H_{ub}$, and $H_{bu}$, and the lowest-order Green's functions $G_{uu}$, $G_{bb}$, $G_{ub}$, and $G_{bu}$ are replaced with their exact counterparts, $\tilde{Q}_{uu}$, $\tilde{Q}_{bb}$, $\cdots$, and $\tilde{G}_{uu}$, $\tilde{G}_{bb}$, $\cdots$, respectively. Under this renormalization procedure on the propagators (spectral and response functions), important turbulent correlation functions are calculated. For the sake of simplicity, hereafter, the tilde denoting an exact propagator will be omitted as $\tilde{Q}_{uu} \to Q_{uu}$, $\tilde{G}_{uu} \to G_{uu}$, etc.

Here we present the final results of the turbulent EMF as
\begin{equation}
	\langle {\bf{u}}' \times {\bf{b}}' \rangle
	= \alpha \langle{\bf{B}}\rangle
	- (\beta + \zeta) \nabla \times \langle{\bf{B}}\rangle
	- (\nabla \zeta) \times \langle{\bf{B}}\rangle
	+ \gamma \left(\langle{\bf{W}}\rangle+2\boldsymbol{\Omega}\right),
		\label{eq:A_emf_result} 
\end{equation}
where transport coefficients $\alpha$, $\beta$, $\zeta$, and $\gamma$ are given as
\begin{equation}
	\alpha 
	= \frac{1}{3} \left[ {
		- I\{ {G_{bb}, H_{uu}} \}
		+ I\{ {G_{uu},H_{bb}} \}
		- I\{ {G_{bu},H_{ub}} \}
		+ I\{ {G_{ub},H_{bu}} \}
	} \right],
		\label{eq:A_alpha_exp}
\end{equation}
\begin{equation}
	\beta 
	= \frac{1}{3} \left[ {
		I\{ {G_{bb}, Q_{uu}} \}
		+ I\{ {G_{uu},Q_{bb}} \}
		- I\{ {G_{bu},Q_{ub}} \}
		- I\{ {G_{ub},Q_{bu}} \}
	} \right],
		\label{eq:A_beta_exp}
\end{equation}
\begin{equation}
	\zeta 
	= \frac{1}{3} \left[ {
		I\{ {G_{bb}, Q_{uu}} \}
		- I\{ {G_{uu},Q_{bb}} \}
		+ I\{ {G_{bu},Q_{ub}} \}
		- I\{ {G_{ub},Q_{bu}} \}
	} \right],
		\label{eq:A_zeta_exp}
\end{equation}
\begin{equation}
	\gamma 
	= \frac{1}{3} \left[ {
		I\{ {G_{bb}, Q_{ub}} \}
		+ I\{ {G_{uu},Q_{bu}} \}
		- I\{ {G_{bu},Q_{uu}} \}
		- I\{ {G_{ub},Q_{bb}} \}
	} \right]
		\label{eq:A_gamma_exp}
\end{equation}
with the abbreviate form of integral
\begin{equation}
	I\{{A,B}\}
	= \int d{\bf{k}} \int_{-\infty}^{\tau}\!\!\! d\tau_1
		A(k;\tau,\tau_1) B(k;\tau,\tau_1).
		\label{eq:A_abbrev_int_form}
\end{equation}

\section{Cross helicity and $\left\langle \mathbf{u}'\cdot\mathbf{j}'\right\rangle $
under FOSA}

In the presence of the Coriolis force and under the 'first-order smoothing approximation' (FOSA), in the Fourier space the linearised equations
take the form
\begin{align}
\left(-\mathrm{i}\omega+\nu k^{2}\right)\hat{u}_{i}(\mathbf{q})+2\Omega\epsilon_{i3j}\hat{u}_{j}(\mathbf{q})= & \hat{f}_{i}(\mathbf{q})+\mathrm{i}\mathbf{k}\cdot\left\langle \mathbf{B}\right\rangle \hat{b}_{i}(\mathbf{q}),\label{eq:u_FOSA}
\end{align}
\begin{align}
\left(-\mathrm{i}\omega+\eta k^{2}\right)\hat{b}_{i}(\mathbf{q})= & \mathrm{i}\mathbf{k}\cdot\left\langle \mathbf{B}\right\rangle \hat{u}_{i}(\mathbf{q}),\label{eq:b_FOSA}
\end{align}
where the the forcing is assumed Gaussian with zero mean, homogeneous,
stationary, and isotropic
\begin{equation}
\left\langle \hat{f}_{i}(\mathbf{k},\omega)\hat{f}_{j}(\mathbf{k}',\omega')\right\rangle =\left[\frac{D_{0}}{k^{3}}P_{ij}(\mathbf{k})+\mathrm{i}\frac{D_{1}}{k^{5}}\epsilon_{ijk}k_{k}\right]\delta(\mathbf{k}+\mathbf{k}')\delta(\omega+\omega'),\label{eq:force_correlations-1}
\end{equation}
and $P_{ij}(\mathbf{k})=\delta_{ij}-k_{i}k_{j}/k^{2}$ is the projection
operator on the plane perpendicular to the wave vector $\mathbf{k}$.
Introducing
\begin{equation}
\gamma_{\nu}=-\mathrm{i}\omega+\nu k^{2},\qquad\gamma_{\eta}=-\mathrm{i}\omega+\eta k^{2},\label{eq:gammas}
\end{equation}
and considering the weak seed field limit defined by
\begin{equation}
\left\langle \mathbf{B}\right\rangle^{2} \ll\left\langle \mathbf{U}\right\rangle^{2} ,\quad\textrm{hence also}\quad\left\langle \mathbf{b}^{\prime2}\right\rangle \ll\left\langle \mathbf{u}^{\prime2}\right\rangle,\label{WSF1}
\end{equation}
the equations reduce to
\begin{equation}
\hat{u}_{i}(\mathbf{q})\approx\mathfrak{G}_{ij}\hat{f}_{j}^{>}(\mathbf{q}),\label{eq:u_and_b_1-1-1}
\end{equation}
\begin{equation}
\hat{b}_{i}(\mathbf{q})\approx\mathrm{i}\frac{\mathbf{k}\cdot\left\langle \mathbf{B}\right\rangle }{\gamma_{\eta}}\mathfrak{G}_{ij}\hat{f}_{j}(\mathbf{q}),\label{eq:b-1}
\end{equation}
where
\begin{equation}
\mathfrak{G}_{ij}=\frac{1}{\gamma_{\nu}^{2}+4\Omega^{2}\frac{k_{z}^{2}}{k^{2}}}\left[\gamma_{\nu}\delta_{ij}-2\Omega\epsilon_{i3j}+2\Omega\frac{k_{i}k_{m}}{k^{2}}\epsilon_{jm3}\right].\label{eq:G_operator}
\end{equation}
The cross-helicity takes the form
\begin{align}
\left\langle h_{ub}\right\rangle = & \left\langle u_{i}^{\prime}(\mathbf{x},t)b_{i}^{\prime}(\mathbf{x},t)\right\rangle \nonumber\\
=&\,\,\mathrm{i}\int\mathrm{d}^{4}q\int\mathrm{d}^{4}q'\mathrm{e}^{\mathrm{i}\left[\left(\mathbf{k}+\mathbf{k}'\right)\cdot\mathbf{x}-\left(\omega+\omega'\right)t\right]}\frac{\mathbf{k}'\cdot\left\langle \mathbf{B}\right\rangle }{\gamma_{\eta}\left(\mathbf{q}'\right)}\mathfrak{G}_{ij}\left(\mathbf{q}\right)\mathfrak{G}_{ik}\left(\mathbf{q}'\right)\left\langle \hat{f}_{j}(\mathbf{q})\hat{f}_{k}(\mathbf{q}')\right\rangle \nonumber \\
= & -\mathrm{i}\left\langle B\right\rangle _{m}\int\mathrm{d}^{4}q\frac{k_{m}}{\gamma_{\eta}\left(-\mathbf{q}\right)}\mathfrak{G}_{ij}\left(\mathbf{q}\right)\mathfrak{G}_{ik}\left(-\mathbf{q}\right)\left[\frac{D_{0}}{k^{3}}P_{jk}(\mathbf{k})+\mathrm{i}\frac{D_{1}}{k^{5}}\epsilon_{jks}k_{s}\right]\nonumber \\
= & -8\Omega\left\langle B\right\rangle _{m}\int\frac{\mathrm{d}k}{k}\int_{-\infty}^{\infty}\mathrm{d}\omega\int_{0}^{2\pi}\mathrm{d}\varphi\int_{-1}^{1}\mathrm{d}X\frac{\omega^{2}D_{1}}{\left(\omega^{2}+\eta^{2}k^{4}\right)\mathcal{F}(\omega,X)}\frac{k_{z}k_{m}}{k^{2}}\nonumber \\
= & -16\pi\left(\left\langle \mathbf{B}\right\rangle \cdot\boldsymbol{\Omega}\right)\int\frac{\mathrm{d}k}{k}\int_{-\infty}^{\infty}\mathrm{d}\omega\int_{-1}^{1}\mathrm{d}X\frac{\omega^{2}X^{2}D_{1}}{\left(\omega^{2}+\eta^{2}k^{4}\right)\mathcal{F}(\omega,X)}\nonumber \\
= & -16\pi D_{1}\mathcal{I}\left(\nu,\eta,\Omega,k_{\ell}\right)\left(\left\langle \mathbf{B}\right\rangle \cdot\boldsymbol{\Omega}\right),\label{eq:hub}
\end{align}
where
\begin{align}
\mathcal{F}(\omega,X)=&\left(\omega^{2}+\nu^{2}k^{4}\right)^{2}-8\Omega^{2}X^{2}\left(\omega^{2}-\nu^{2}k^{4}\right)+16\Omega^{4}X^{4}\nonumber\\
=&\left(\omega^{2}-4\Omega^{2}X^{2}\right)^{2}+2\omega^{2}\nu^{2}k^{4}+8\nu^{2}k^{4}\Omega^{2}X^{2}+\nu^{4}k^{8}>0,
\label{Denominator}
\end{align}
and
\begin{equation}
D_{1}\sim\left\langle \mathbf{f}\cdot\nabla\times\mathbf{f}\right\rangle ,\qquad\mathcal{I}\left(\nu,\eta,\Omega,k_{\ell}\right)>0.\label{eq:D1}
\end{equation}
On the other hand for the scalar quantity $\left\langle s_{uj}\right\rangle =\left\langle \mathbf{u}'\cdot\mathbf{j}'\right\rangle $
this approach yields
\begin{align}
\left\langle s_{uj}\right\rangle = & \left\langle u_{i}^{\prime}(\mathbf{x},t)j_{i}^{\prime}(\mathbf{x},t)\right\rangle \nonumber\\
=&-\epsilon_{irt}\left\langle B\right\rangle _{m}\int\mathrm{d}^{4}q\int\mathrm{d}^{4}q'\mathrm{e}^{\mathrm{i}\left[\left(\mathbf{k}+\mathbf{k}'\right)\cdot\mathbf{x}-\left(\omega+\omega'\right)t\right]}\frac{k_{r}^{\prime}k_{m}^{\prime}}{\gamma_{\eta}\left(\mathbf{q}'\right)}\mathfrak{G}_{ij}\left(\mathbf{q}\right)\mathfrak{G}_{tk}\left(\mathbf{q}'\right)\left\langle \hat{f}_{j}(\mathbf{q})\hat{f}_{k}(\mathbf{q}')\right\rangle \nonumber \\
= & -\epsilon_{irt}\left\langle B\right\rangle _{m}\int\mathrm{d}^{4}q\frac{k_{r}k_{m}}{\gamma_{\eta}\left(-\mathbf{q}\right)}\mathfrak{G}_{ij}\left(\mathbf{q}\right)\mathfrak{G}_{ik}\left(-\mathbf{q}\right)\left[\frac{D_{0}}{k^{3}}P_{jk}(\mathbf{k})+\mathrm{i}\frac{D_{1}}{k^{5}}\epsilon_{jks}k_{s}\right]\nonumber \\
= & \,\,8\Omega\left\langle B\right\rangle _{m}\int k\mathrm{d}k\int_{-\infty}^{\infty}\mathrm{d}\omega\int_{0}^{2\pi}\mathrm{d}\varphi\int_{-1}^{1}\mathrm{d}X\frac{\omega^{2}D_{0}}{\left(\omega^{2}+\eta^{2}k^{4}\right)\mathcal{F}(\omega,X)}\frac{k_{z}k_{m}}{k^{2}}\nonumber \\
= & \,\,16\pi\left(\left\langle \mathbf{B}\right\rangle \cdot\boldsymbol{\Omega}\right)\int k\mathrm{d}k\int_{-\infty}^{\infty}\mathrm{d}\omega\int_{-1}^{1}\mathrm{d}X\frac{\omega^{2}X^{2}D_{0}}{\left(\omega^{2}+\eta^{2}k^{4}\right)\mathcal{F}(\omega,X)}\nonumber \\
= & \,\,16\pi D_{0}\widetilde{\mathcal{I}}\left(\nu,\eta,\Omega,k_{\ell}\right)\left(\left\langle \mathbf{B}\right\rangle \cdot\boldsymbol{\Omega}\right),\label{eq:uj}
\end{align}
where
\begin{equation}
D_{0}\sim\left\langle \mathbf{f}^{2}\right\rangle ,\qquad\widetilde{\mathcal{I}}\left(\nu,\eta,\Omega,k_{\ell}\right)>0.\label{eq:D0}
\end{equation}
In the above we have used
\begin{equation}
\int\frac{k_{j}}{k}f\left(\cos^{2}\theta\right)\mathrm{d}\mathring{\varOmega}=0,\qquad\int\frac{k_{i}k_{j}k_{k}}{k^{3}}f\left(\cos^{2}\theta\right)\mathrm{d}\mathring{\varOmega}=0,\label{eq:delta_formula_1}
\end{equation}
\begin{equation}
\int\frac{k_{j}k_{n}}{k^{2}}f\left(\cos^{2}\theta\right)\mathrm{d}\mathring{\varOmega}=\pi\int_{-1}^{1}f(X^{2})\left\{ \delta_{jn}\left(1-X^{2}\right)+\delta_{j3}\delta_{n3}\left(3X^{2}-1\right)\right\} \mathrm{d}X.\label{eq:delta_formula_2}
\end{equation}
where $\mathring{\varOmega}$ denotes the solid angle and the spherical
coordinates $(k,\theta,\varphi)$ have been used (with a substitution
$X=\cos\theta$). Furthermore, in a similar way we can calculate the kinetic helicity and turbulent energy
\begin{align}
\left\langle h_{kin}\right\rangle = & \left\langle u_{i}(\mathbf{x},t)w_{i}(\mathbf{x},t)\right\rangle\nonumber\\
= &\,\, \mathrm{i}\epsilon_{ijk}\int\mathrm{d}^{4}q\int\mathrm{d}^{4}q'k_{j}^{\prime}\mathfrak{G}_{in}\left(\mathbf{q}\right)\mathfrak{G}_{km}\left(\mathbf{q}'\right)\left\langle \hat{f}_{n}(\mathbf{q})\hat{f}_{m}(\mathbf{q}')\right\rangle \mathrm{e}^{\mathrm{i}\left[\left(\mathbf{k}+\mathbf{k}'\right)\cdot\mathbf{x}-\left(\omega+\omega'\right)t\right]}\nonumber \\
= & -\mathrm{i}\epsilon_{ijk}\int\mathrm{d}^{4}qk_{j}\mathfrak{G}_{in}\left(\mathbf{q}\right)\mathfrak{G}_{km}\left(-\mathbf{q}\right)\left[\frac{D_{0}}{k^{3}}P_{nm}(\mathbf{k})+\mathrm{i}\frac{D_{1}}{k^{5}}\epsilon_{nmp}k_{p}\right]\nonumber \\
= & -\int\frac{\mathrm{d}k}{k}\int_{-\infty}^{\infty}\mathrm{d}\omega\int_{0}^{2\pi}\mathrm{d}\varphi\int_{-1}^{1}\mathrm{d}X\frac{2D_{1}\left(\omega^{2}+\nu^{2}k^{4}+4\Omega^{2}X^{2}\right)}{\left(\omega^{2}+\nu^{2}k^{4}\right)^{2}-8\Omega^{2}X^{2}\left(\omega^{2}-\nu^{2}k^{4}\right)+16\Omega^{4}X^{4}}\nonumber \\
= & -4\pi D_{1}\mathcal{I}_{u^{2}}\left(\nu,\Omega,k_{\ell}\right)\label{eq:hk_gen}
\end{align}
\begin{align}
\left\langle \mathbf{u}^{\prime2}\right\rangle = & \left\langle u_{i}(\mathbf{x},t)u_{i}(\mathbf{x},t)\right\rangle \nonumber\\
=&\int\mathrm{d}^{4}q\int\mathrm{d}^{4}q'\mathfrak{G}_{in}\left(\mathbf{q}\right)\mathfrak{G}_{im}\left(\mathbf{q}'\right)\left\langle \hat{f}_{n}(\mathbf{q})\hat{f}_{m}(\mathbf{q}')\right\rangle \mathrm{e}^{\mathrm{i}\left[\left(\mathbf{k}+\mathbf{k}'\right)\cdot\mathbf{x}-\left(\omega+\omega'\right)t\right]}\nonumber \\
= & \int\mathrm{d}^{4}q\mathfrak{G}_{in}\left(\mathbf{q}\right)\mathfrak{G}_{im}\left(-\mathbf{q}\right)\left[\frac{D_{0}}{k^{3}}P_{nm}(\mathbf{k})+\mathrm{i}\frac{D_{1}}{k^{5}}\epsilon_{nmp}k_{p}\right]\nonumber \\
= & \,\,2\int\frac{\mathrm{d}k}{k}\int_{-\infty}^{\infty}\mathrm{d}\omega\int_{0}^{2\pi}\mathrm{d}\varphi\int_{-1}^{1}\mathrm{d}X\frac{D_{0}\left(\omega^{2}+\nu^{2}k^{4}+4\Omega^{2}X^{2}\right)}{\left(\omega^{2}+\nu^{2}k^{4}\right)^{2}-8\Omega^{2}X^{2}\left(\omega^{2}-\nu^{2}k^{4}\right)+16\Omega^{4}X^{4}}\nonumber \\
= & \,\,4\pi D_{0}\mathcal{I}_{u^{2}}\left(\nu,\Omega,k_{\ell}\right).\label{eq:u2}
\end{align}
Note, that in the weak seed field limit (\ref{WSF1}) the turbulent energy reduces to $\left\langle \mathbf{u}^{\prime2}\right\rangle +\left\langle \mathbf{b}^{\prime2}\right\rangle \approx\left\langle \mathbf{u}^{\prime2}\right\rangle
$. We can now utilize the above results to show
\begin{align}
\left\langle \mathbf{u}^{\prime}\cdot\mathbf{j}^{\prime}\right\rangle = & \left(\frac{\mathcal{I}_{uj}\left(\nu,\eta,\Omega,k_{\ell}\right)}{\mathcal{I}_{ub}\left(\nu,\eta,\Omega,k_{\ell}\right)}\frac{D_{0}^{2}}{D_{1}^{2}}\right)\frac{\left\langle \mathbf{u}^{\prime}\cdot\mathbf{b}^{\prime}\right\rangle }{\left\langle \mathbf{u}^{\prime2}\right\rangle }\left\langle \mathbf{u}^{\prime}\cdot\mathbf{w}^{\prime}\right\rangle \nonumber\\
\sim & \,\, \frac{\left\langle \mathbf{u}^{\prime}\cdot\mathbf{b}^{\prime}\right\rangle }{\left\langle \mathbf{u}^{\prime2}\right\rangle }\left\langle \mathbf{u}^{\prime}\cdot\mathbf{w}^{\prime}\right\rangle  \label{uj_vs_uo}
\end{align}

\section{Evolution equations for $\langle \mathbf{u}'\cdot\mathbf{b}'\rangle$ and $\langle \mathbf{u}'\cdot\mathbf{j}'\rangle$}

Utilizing the evolution equations
\begin{align}
\frac{D\mathbf{u}'}{Dt}=-\nabla\Pi'-2\boldsymbol{\Omega}\times\mathbf{u}'-\left(\mathbf{u}'\cdot\nabla\right)\left\langle \mathbf{U}\right\rangle +\left(\left\langle \mathbf{B}\right\rangle \cdot\nabla\right)\mathbf{b}'+\left(\mathbf{b}'\cdot\nabla\right)\left\langle \mathbf{B}\right\rangle +\nu\nabla^{2}\mathbf{u}' & \nonumber \\
 & \hspace{-6cm}+\nabla\cdot\left(\mathbf{b}'\mathbf{b}'\right)+\nabla\cdot\left(\left\langle \mathbf{u}'\mathbf{u}'\right\rangle -\left\langle \mathbf{b}'\mathbf{b}'\right\rangle \right),\label{eq:Fluct_u_eq_1}
\end{align}
\begin{equation}
\frac{D\mathbf{b}'}{Dt}=\left(\left\langle \mathbf{B}\right\rangle \cdot\nabla\right)\mathbf{u}'-\left(\mathbf{u}'\cdot\nabla\right)\left\langle \mathbf{B}\right\rangle +\left(\mathbf{b}'\cdot\nabla\right)\left\langle \mathbf{U}\right\rangle +\eta\nabla^{2}\mathbf{b}'+\left(\mathbf{b}'\cdot\nabla\right)\mathbf{u}'-\nabla\times\boldsymbol{\mathcal{E}},\label{eq:Fluct_b_eq_1}
\end{equation}
where
\begin{equation}
\frac{D}{Dt} = \frac{\partial}{\partial t} + \left(\langle\mathbf{U}\rangle+\mathbf{u}'\right)\cdot\nabla,\label{dbydt}
\end{equation}
we arrive at
\begin{align}
\frac{D}{Dt}\left\langle \mathbf{u}^{\prime}\cdot\mathbf{b}^{\prime}\right\rangle
= & -\boldsymbol{\mathcal{E}}\cdot\left(\left\langle \mathbf{W}\right\rangle +2\boldsymbol{\Omega}\right)-\left\langle u_{i}^{\prime}u_{j}^{\prime}-b_{i}^{\prime}b_{j}^{\prime}\right\rangle \partial_{j}\left\langle B\right\rangle _{i}\nonumber \\
 & +\nabla\cdot\left[\left\langle \left(-\Pi'+\frac{\mathbf{u}^{\prime2}+\mathbf{b}^{\prime2}}{2}\right)\mathbf{b}'\right\rangle +\left\langle \frac{\mathbf{u}^{\prime2}+\mathbf{b}^{\prime2}}{2}\right\rangle \left\langle \mathbf{B}\right\rangle \right]\nonumber\\
 & -\mu_{0}\left(\nu+\eta\right)\left\langle \mathbf{w}'\cdot\mathbf{j}^{\prime}\right\rangle, \label{ub}
\end{align}
and
\begin{align}
\frac{D}{Dt}\left\langle \mathbf{u}^{\prime}\cdot\mathbf{j}^{\prime}\right\rangle
= & -\left\langle \mathbf{u}'\times\mathbf{j}^{\prime}\right\rangle \cdot\left(\left\langle \mathbf{W}\right\rangle +2\boldsymbol{\Omega}\right)-\left\langle w_{i}^{\prime}u_{j}^{\prime}-j_{i}^{\prime}b_{j}^{\prime}\right\rangle \partial_{j}\left\langle B\right\rangle _{i}-\partial_{j}\left\langle U\right\rangle _{m}\left\langle \epsilon_{ijk}u_{i}^{\prime}\partial_{m}b_{k}^{\prime}\right\rangle \nonumber\\
 &  +\left\langle \left[\left(\left\langle \mathbf{B}\right\rangle +\mathbf{b}'\right)\cdot\nabla\right]\mathbf{b}'\cdot\mathbf{j}^{\prime}+\left[\left(\left\langle \mathbf{B}\right\rangle +\mathbf{b}'\right)\cdot\nabla\right]\mathbf{u}^{\prime}\cdot\mathbf{w}^{\prime}-u_{i}^{\prime}\epsilon_{ijk}\partial_{j}u_{m}^{\prime}\partial_{m}b_{k}^{\prime}\right\rangle \nonumber\\
 & -\nabla\cdot\left\langle \Pi'\mathbf{j}^{\prime}\right\rangle-\left(\nu-\eta\right)\left\langle \mathbf{w}^{\prime}\cdot\nabla^{2}\mathbf{b}^{\prime}\right\rangle, \label{uj} 
\end{align}
where in the last equation, apart from no-slip boundary conditions, we have also assumed vanishing of the helical quantity $\left\langle \mathbf{w}'\cdot\mathbf{j}^{\prime}\right\rangle$ at the boundaries.

\section{Basic equations used in the compressible case}

In the numerical simulations, instead of equations (\ref{eq:NS}--c),
we solve the following set of equations for a compressible isothermal
gas with constant sound speed $c_{\rm s}$ for $\mathbf{U}$, $\rho$, and the
magnetic vector potential $\mathbf{A}$:
\begin{subequations}
\begin{equation}
\frac{\partial\mathbf{U}}{\partial t}+\left(\mathbf{U}\cdot\nabla\right)\mathbf{U}=-c_{\rm s}^2\nabla\ln\rho-2\boldsymbol{\Omega}\times\mathbf{U}
+\frac{1}{\rho} \mathbf{J}\times\mathbf{B}-\nu\mathbf{Q}+\mathbf{g}+\mathbf{f}
,\label{eq:compNS}
\end{equation}
\begin{equation}
\frac{\partial\rho}{\partial t}=-\nabla\cdot(\rho\mathbf{U}),\label{eq:compDIVS}
\end{equation}
\end{subequations}
\begin{equation}
\frac{\partial\mathbf{A}}{\partial t}=\mathbf{U}\times\mathbf{B}-\eta\mu_0\mathbf{J}
,\label{eq:compIND}
\end{equation}
where
\begin{equation}
\mathbf{Q}=-\nabla^{2}\mathbf{U}-\frac{1}{3}\nabla\nabla\cdot\mathbf{U}-\mbox{\boldmath ${\sf S}$}\nabla\ln\rho,
\end{equation}
\begin{equation}
\mu_0\mathbf{J}=-\nabla^{2}\mathbf{A}+\nabla\nabla\cdot\mathbf{A},
\end{equation}
\begin{equation}
\mathbf{B}=\mathbf{B}_0+\nabla\times\mathbf{A},
\end{equation}
and
\begin{equation}
\mathsf{S}_{ij}=\frac{1}{2}(\partial_i U_j+\partial_j U_i)-\frac{1}{3}\delta_{ij}\nabla\cdot\mathbf{U}
\end{equation}
are the components of the traceless rate-of-strain tensor and $\mathbf{f}$
is a random forcing function consisting of plane unpolarized waves
with typical wavenumber $k_{\rm f}$ and an amplitude such that
$u_{\rm rms}/c_{\rm s}\approx0.1$; see table~1.
Here, $\boldsymbol{\Omega}=(0,0,\Omega)$ is the angular velocity,
$\mathbf{g}=(0,0,-g)$ is gravity,
$\mathbf{B}_0=(0,0,B_0)$ is the imposed magnetic field,
$\eta$ is the magnetic diffusivity, and $\nu$ is the kinematic viscosity,
whose value is such that $u_{\rm rms}/\nu k_1\approx1000$.
A resolution of $N^3=256^3$ mesh points is then sufficient.
Since we chose $k_{\rm f}/k_1=30$, we have for the Reynolds number
$\mbox{Re}\equiv u_{\rm rms}/\nu k_{\rm f}\approx30$.
For the magnetic Prandtl number we chose, as in Jabbari et al. (2014)
the value $\mbox{Pr}_M\equiv\nu/\eta=0.5$, so the magnetic Reynolds number is
$\mbox{Re}_M\equiv u_{\rm rms}/\eta k_{\rm f}\approx15$.
The equilibrium stratification is given by $\ln(\rho/\rho_0)=-z/H_\rho$,
where $H_\rho=c_{\rm s}^2/g$ is the density scale height.

\section{Results for Runs~B--E}

In figure~2, we present the results for Runs~B and E with a stronger
magnetic field: $B=0.1\,c_{\rm s}\sqrt{\mu_0\bar{\rho}}$, and two values
of gravity $g=1\, c_{\rm s}^2 k_1$ and $g=0.5\, c_{\rm s}^2 k_1$.
Finally, in figure~3, we present the results for Run~C and D with weaker
magnetic field $B=0.01\,c_{\rm s}\sqrt{\mu_0\bar{\rho}}$, and two values
of gravity: $g=0.5\, c_{\rm s}^2 k_1$ and $g=2\, c_{\rm s}^2 k_1$.

\begin{figure}
\begin{centering}
\includegraphics[width=0.49\linewidth]{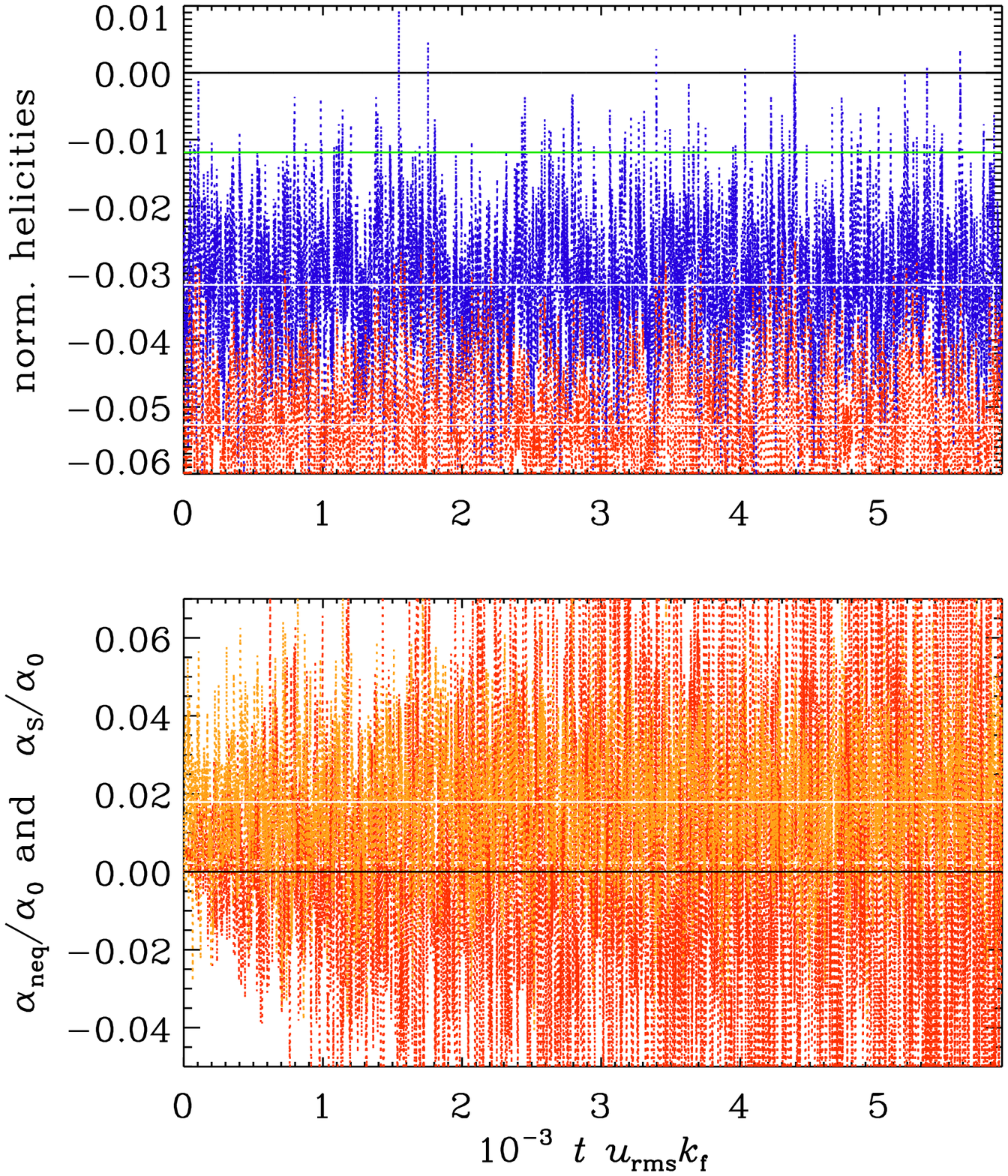}
\includegraphics[width=0.49\linewidth]{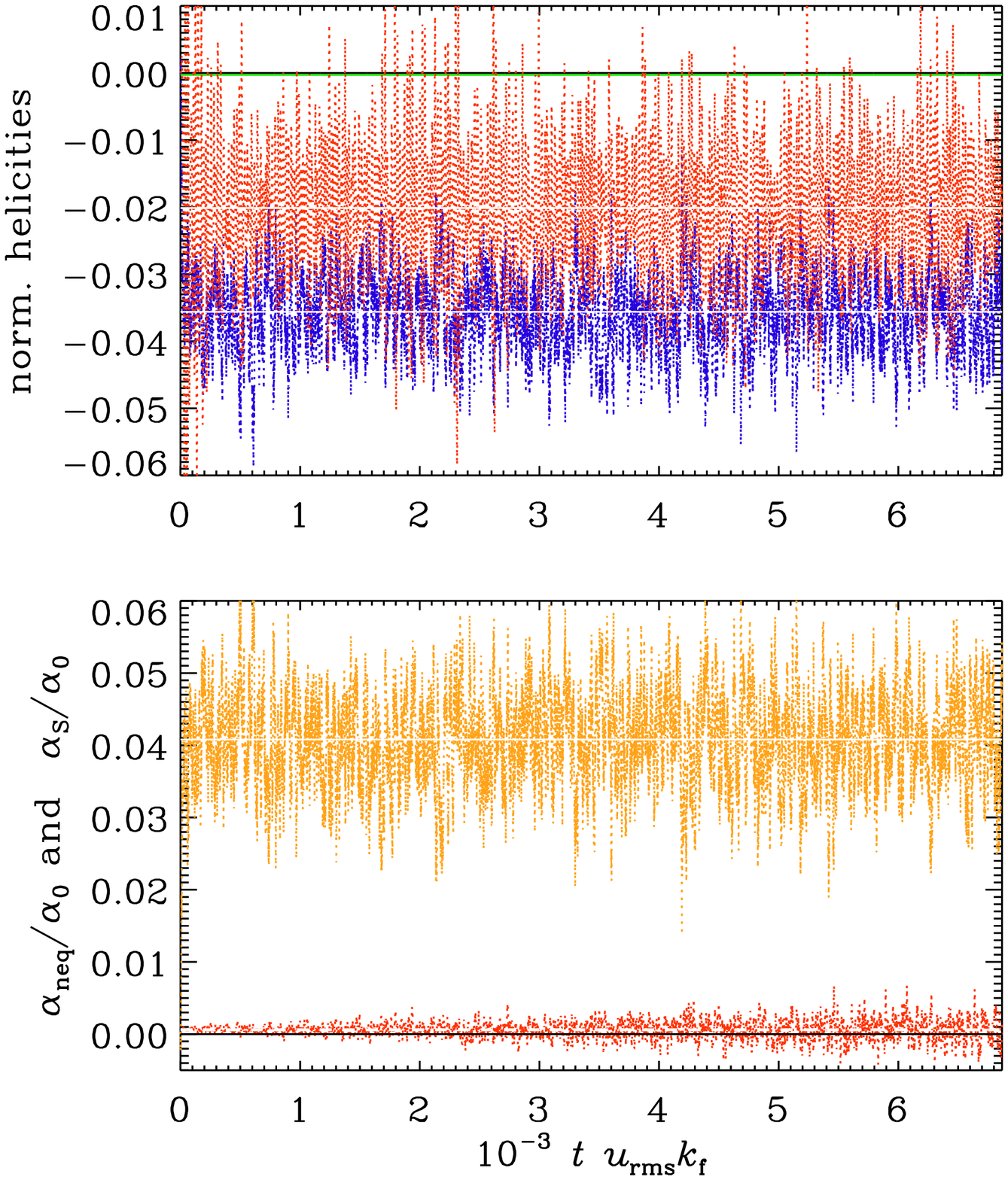}
\par\end{centering}
\caption{\label{fig:b0-1}Same as figure \ref{fig:b0-01}, but for Run~B and E with a stronger magnetic field: $B=0.1\,c_{\rm s}\sqrt{\mu_0\bar{\rho}}$,
and two values of gravity $g=1\, c_{\rm s}^2 k_1$ and $g=0.5\, c_{\rm s}^2 k_1$.}
\end{figure}
\begin{figure}
\begin{centering}
\includegraphics[width=0.49\linewidth]{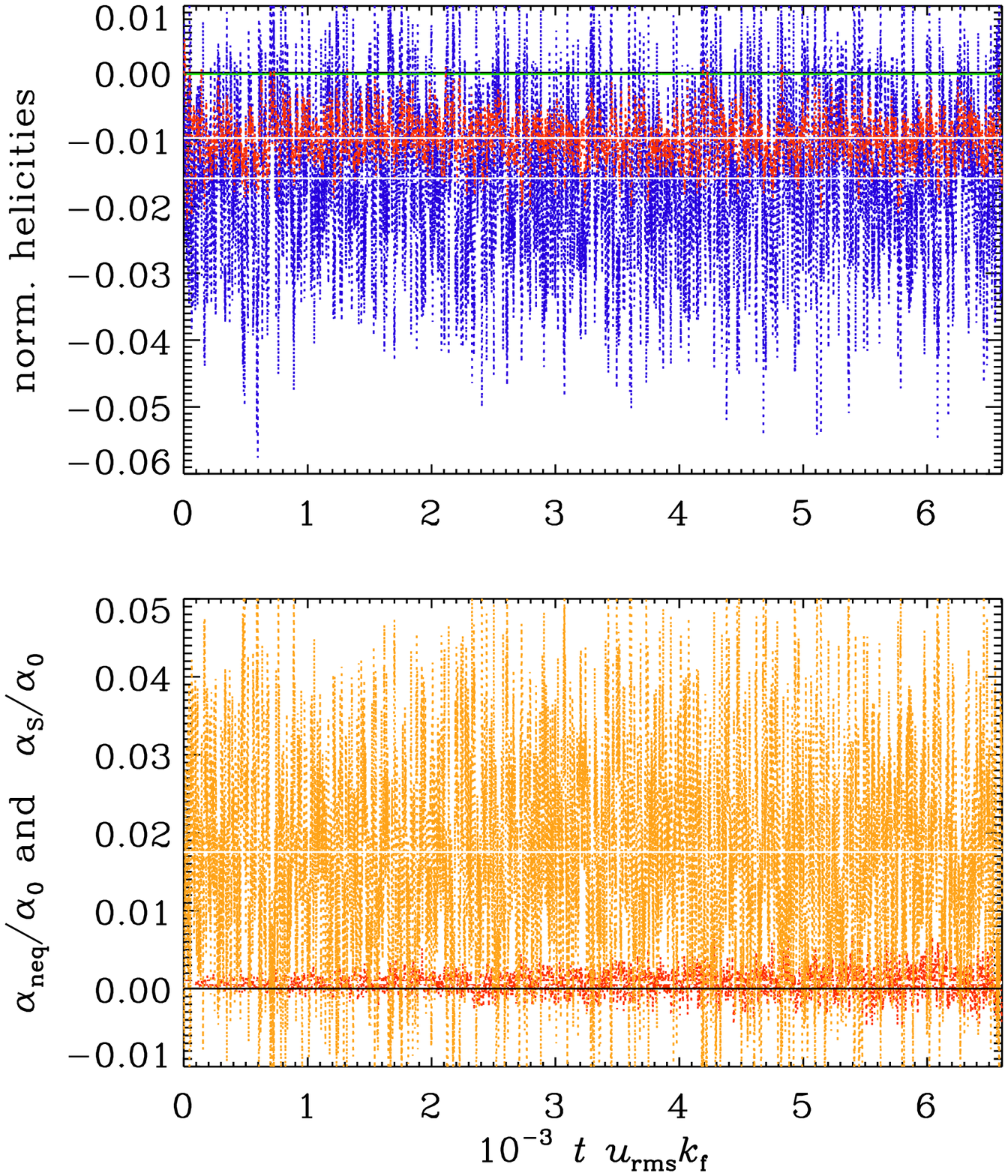}
\includegraphics[width=0.49\linewidth]{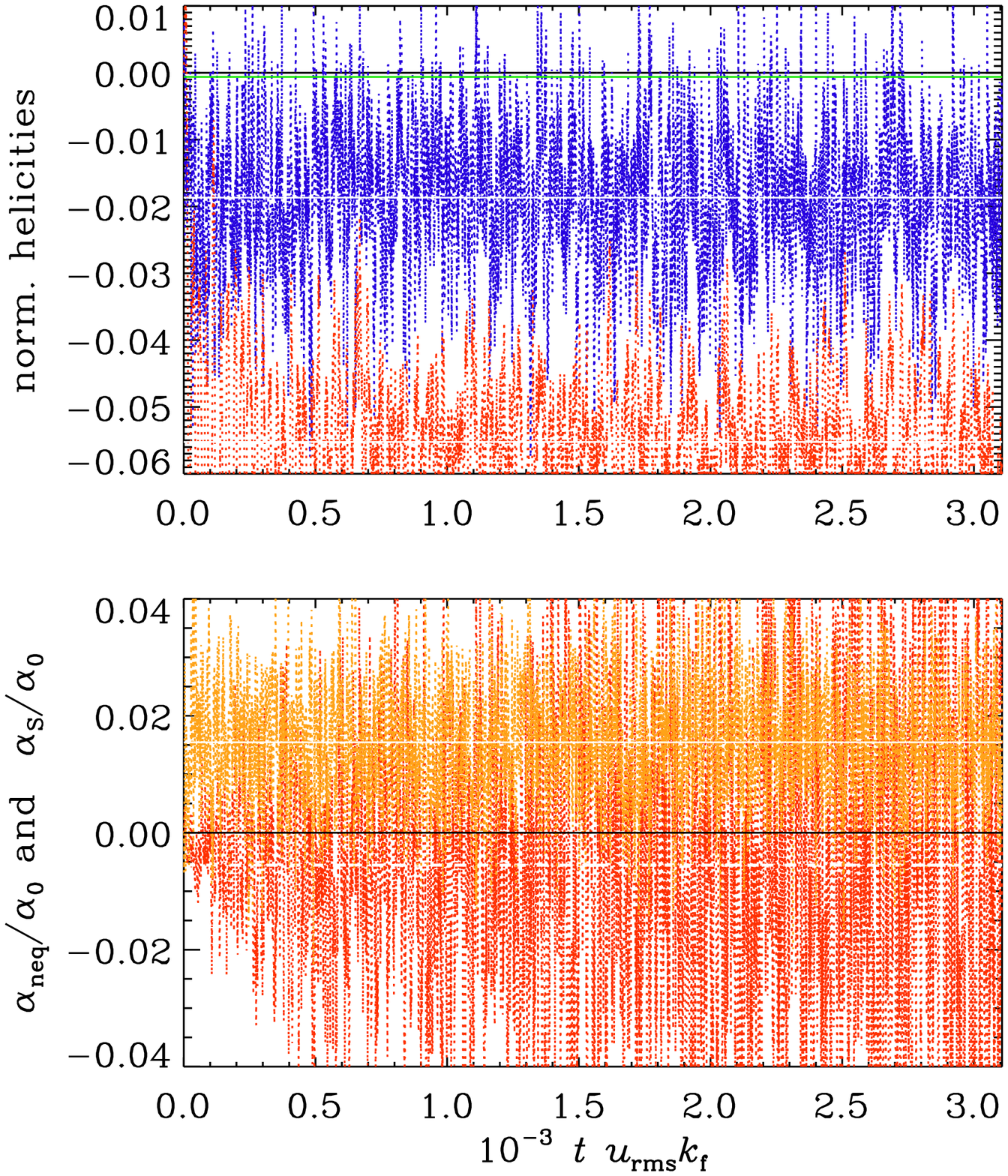}
\par\end{centering}
\caption{\label{fig:b0-01_diff_g}Same as figure \ref{fig:b0-01}, but for Run~C and D with weaker magnetic field $B=0.01\,c_{\rm s}\sqrt{\mu_0\bar{\rho}}$, and two values of gravity: $g=0.5\, c_{\rm s}^2 k_1$ and $g=2\, c_{\rm s}^2 k_1$.}
\end{figure}
%


\bibliographystyle{jpp}


\end{document}